\documentclass[conference,compsoc]{IEEEtran}
\IEEEoverridecommandlockouts
\usepackage{float}
\usepackage{placeins}
\usepackage{hyperref}
\usepackage{caption}
\usepackage{subcaption}
\usepackage{graphicx}
\usepackage{tabularx}
\usepackage{verbatim, color}
\usepackage{listings}
\definecolor{dkgreen}{rgb}{0,0.6,0}
\definecolor{gray}{rgb}{0.5,0.5,0.5}
\definecolor{mauve}{rgb}{0.58,0,0.82}

\usepackage[most]{tcolorbox}
\usepackage{filecontents}
\usepackage{CJK}
\usepackage{soul}
\usepackage{blindtext}
\newcounter{observation}
\usepackage{multirow, makecell}
\usepackage{cite}
\usepackage{amsmath,amssymb,amsfonts}
\usepackage{algorithmic}
\usepackage{graphicx}
\usepackage{textcomp}
\usepackage{xcolor}
\def\BibTeX{{\rm B\kern-.05em{\sc i\kern-.025em b}\kern-.08em
    T\kern-.1667em\lower.7ex\hbox{E}\kern-.125emX}}

\begin{document}

\date{}
\newcommand{\tanusree}[1]{{\color{cyan} \textbf{(Tanusree: #1)}}}
\newcommand{\yujin}[1]{{\color{olive} \textbf{(Yujin: #1)}}}
\newcommand{\korn}[1]{{\color{orange} \textbf{(Korn: #1)}}}
\newcommand{\dawn}[1]{{\color{red}\textbf{(Dawn: #1)}}}
\newcommand{\yang}[1]{{\color{blue} \textbf{(Yang: #1)}}}
\newcommand{\anote}[1]{{\color{magenta} \textbf{(AM: #1)}}}
\newcommand{\update}[1]{{\color{green} \textbf{(update: #1)}}}
\newcommand{\fixme}[1]{{\color{red} \textbf{(#1)}}}
\title{
Unpacking How Decentralized Autonomous Organizations (DAOs) Work in Practice \vspace{-4mm}} 



\author{


{\rm Tanusree Sharma\textsuperscript{*}, Yujin Kwon\textsuperscript{\dag}, Kornrapat Pongmala\textsuperscript{\dag}, Henry Wang\textsuperscript{*}, Andrew Miller\textsuperscript{*}, Dawn Song\textsuperscript{\dag}, Yang Wang\textsuperscript{*}}\\
 {\rm \textsuperscript{*}University of Illinois at Urbana-Champaign 
  \hspace{0.05in}
  \textsuperscript{\dag}University of California, Berkeley}
  \\

}

\maketitle


\begin{abstract}
Decentralized Autonomous Organizations (DAOs) have emerged as a novel way to coordinate a group of (pseudonymous) entities towards a shared vision (e.g., promoting sustainability), utilizing self-executing smart contracts on blockchains to support decentralized governance and decision-making. In just a few years, over 4,000 DAOs have been launched in various domains, such as investment, education, health, and research. Despite such rapid growth and diversity, it is unclear how these DAOs actually work in practice and to what extent they are effective in achieving their goals. 
Given this, we aim to unpack how (well) DAOs work in practice. 
We conducted an in-depth analysis of a diverse set of 10 DAOs of various categories and smart contracts, leveraging on-chain (e.g., voting results) and off-chain data (e.g., community discussions) as well as our interviews with DAO organizers/members. Specifically, we defined metrics to characterize key aspects of DAOs, such as the degrees of decentralization and autonomy. We observed CompoundDAO, AssangeDAO, Bankless, and Krausehouse
having poor decentralization in voting, while decentralization has
improved over time for one-person-one-vote DAOs (e.g., Proof of Humanity). Moreover, the degree of autonomy varies among DAOs, with some (e.g., Compound and Krausehouse) relying more on third parties than others. 
Lastly, we offer a set of design implications for future DAO systems based on our findings.

\end{abstract}

\vspace{-2mm}
\section{Introduction}
\vspace{-2mm}

As of 2022, there have been over 4,000 decentralized autonomous organizations (DAOs) created with the goal to democratize the management structure for decentralized protocols and open-source projects~\cite{messari} utilizing smart contracts (i.e., computer programs encoded with specific rules) to govern their operations~\cite{dwivedi2021legally}. For example, in ConstitutionDAO, where the rules and decision-making processes are encoded into a blockchain, allowing the terms of agreement between buyers and sellers to be directly written into lines of code. UkraineDAO is another example which was built to launch a crowdfunding campaign to mint and sell non-fungible tokens (NFTs) of a Ukrainian flag to raise funds for civilian and military purposes~\cite{esteves2023potential,chaisse2022tokenised}. One of the key functions of a DAO is group decision-making, which is often conducted via a series of proposals where members vote with the DAO's governance tokens~\cite{coindesk}. The market value of such governance tokens typically signifies the relative influence of stakeholders within the DAO.
Despite challenges faced by early DAOs, there are governance benefits to this structure, e.g., community-driven consensus, execution of predefined rules, transparency of the decision-making process~\cite{hassan2021decentralized, hackernoon}.

However, the governance of DAOs is a complex issue; it is not yet clear to what degree they attain decentralization and autonomy, even though the name of DAO implies the two. Specifically, the influence of community involvement, decision-making authority, and positioning of DAO members in proposals and voting mechanisms on the level of decentralization remain unclear~\cite{zachariadis2019governance}. Furthermore, the extent to which DAOs operate without undue external influence (from corporate competitors, regulators, or third parties in general) in practice has not yet been explored. Therefore, systematic exploration is required in terms of procedural tedium\footnote{In the context of smart contract, while it aims to streamline protocol execution in the digital era, but still relies on tedious human involvement}, 
 structural rigidity, community interaction, and voting manipulation through on-chain and off-chain data~\cite{zwitter2020decentralized, zwitter2020governance}. In this paper, we aim to discover how concretely  DAOs work in practice (focusing on decentralization and autonomy). We address the following research questions.

\noindent \textbf{RQ1: What are the current perceptions \& practices of DAO participants in fulfilling DAOs' vision? 
} 
We conducted one-on-one interviews with 10 DAO participants and two focus groups (each with a dozen of DAO participants). 
Our results reveal their nuanced perceptions and practices, e.g., \textit{``being decentralized is not the sole purpose''} and 
why DAO participants often voted in a similar way was due to the influence of ``whales'' (i.e., those with a huge amount of governance tokens), trust in key opinion leaders, and peer pressure. 
%
Our DAO participants varied in their definitions of DAO autonomy. While some viewed autonomy as \textit{``less human intervention, more tool use,''} others felt it is more about {\it without undue external influence}. 
We referred to the interview findings to establish metrics used in our follow-up analysis to explore RQ2 and RQ3.


\noindent \textbf{RQ2: How decentralized is a DAO?} 
We investigate the decentralization degree of DAOs by analyzing on-chain and off-chain data.
We find that token holdings have a significant positive correlation with proposal success in KrauseHouse, CompoundDAO, and dxDAO while for BitDAO, BanklessDAO, and AssangeDAO, there is no significant correlation, likely due to the presence of predefined whitelisted authors, community elected proposal authors, etc. CompoundDAO, AssangeDAO, Bankless, and Krausehouse exhibit high polarization in voting patterns indicating poor decentralization while Proof of Humanity (PoH), Meta Gamma Delta, and Moloch have a more equitable power distribution. Moreover, we find that members with more governance tokens tend to vote more actively, which deteriorates decentralization.   
Finally, the decentralization level has been gradually aggravated in CompoundDAO, BitDAO, and BanklessDAO, which have governance based on token capital, while it has improved over time in the one-person-one-vote DAO, PoH. 


\noindent\textbf{
RQ3: How autonomous is a DAO?
}
Based on inputs from our interviews and focus groups, we considered two aspects of DAO autonomy: (1) less human intervention, more tool usage; and (2) without undue external influence.
%
Our analysis revealed variations among DAOs to achieve arbitrary transaction execution, with on-chain proposal submission being the most autonomous via code modules. 
We also observed most DAOs require 3rd-party services to operate. 

\noindent{\bf Main contributions.} Our work makes the following contributions: 
(1) We offer novel insights into the perceptions and practices regarding the degree of decentralization and autonomy in fulfilling the visions of DAOs by interviewing DAO participants.
(2) Through our empirical analysis of on-chain and off-chain voting/proposal systems, as well as assessments of corresponding smart contracts, we shed light on the level of decentralization and autonomy in DAOs using key metrics, some of which we learned from the DAO participant interviews. 
(3) Based on our empirical results, we discuss design implications for governance models of DAOs and DAO tooling such as DAO proposal transparency tools. 

\vspace{-2mm}
\section{Related Work}
\label{related_work}
\vspace{-2mm}
\subsection{History of DAOs}
\vspace{-2mm}
The concept of a DAO has existed since the mid-2010s~\cite{chohan2017decentralized}. \emph{The DAO} was originally designed as an investor-driven venture capital fund that relied on voting by investors to disburse funds to proposals submitted by contractors and vetted by curators~\cite{mehar2019understanding}. It operates as a transparent and democratically structured virtual platform, without physical addresses or formal managerial roles. Despite its potential of launching one of the largest crowdfunded campaigns ever seen~\cite{liu2021technology}, it was immediately hacked and drained of \$50 million in cryptocurrency~\cite{dhillon2017dao}, highlighting a mismatch between the system's openness and the potential for nefarious actions~\cite{liu2021technology, morrison2020dao, dupont2017experiments}. Yet, this should not conflate the broader category of smart contract based similar technologies, such as Dash governance~\cite{mosley2022towards}, Digix.io\footnote{Digix.io is a smart-asset gold-focused coin that seeks to match its value with the price of physical gold.}, Augur\footnote{Augur centers around the prediction markets and betting arenas where financial options and insurance markets can be developed.}, Uniswap\footnote{Uniswap is a crypto exchange based on smart contracts}. Many of these focused on blockchain-based assets and digital variants of existing socioeconomic instruments such as insurance, exchange markets, and social media~\cite{dupont2017experiments}. While some researchers argue that DAOs were initially limited to private capital allocation~\cite{chohan2017decentralized, trisetyarso2019crypto}, there is a growing trend to use DAOs in high-value data, and reputational-based systems~\cite{myeong2019administrative, barbosa2018cryptocurrencies,chohan2017decentralized}. The deterministic and non-probabilistic nature of smart contracts can be adapted~\cite{chatterjee2019probabilistic} towards these new paradigms based on the refinement of programming logic of organizational rules~\cite{ciatto2018blockchain}. 

Although it is difficult to describe DAOs as an organizational type~\cite{tse2020decentralised, trisetyarso2019crypto}, the conceptual basis of a DAO is its underlying ability within blockchain to provide a secure digital ledger that tracks financial interactions across the internet, bolstered against forgery by trusted timestamping and a distributed database. Users and practitioners who argue for decentralized and autonomous technologically-driven solutions have supported this idea to reflect and facilitate virtualization tendencies~\cite{ellul2019blockchain}. 
Unlike traditional capitalist organizations with undemocratic decision-making processes, where power is concentrated among boards, management, and shareholders, according to Marxist theory~\cite{bowles2012democracy}, DAOs offer a decentralized alternative, allowing for democratic decision-making through consensus protocols~\cite{dupont2017experiments}.
However, there has been the ever-continuing discussion around relegating management to pre-programmed rules, where transparency and efficiency are marked positives, but unstable security and low flexibility are the disadvantages~\cite{chohan2017decentralized}.

\vspace{-2mm}
\subsection{DAO in Context of Co-ordination
}
\vspace{-2mm}
DAOs enable individuals to coordinate and govern through self-executing rules on blockchain, without centralized control~\cite{buterin2014next}. While the primary objective of a DAO is to replicate traditional organizations with distributed decision-making by using blockchain technology, the precise definition of a DAO is currently contested~\cite{dupont2017experiments}. Some argue that all cryptocurrencies can be considered DAOs, while others propose that decentralized and autonomous governance structures could be considered the original DAOs~\cite{morrison2020dao}. As such, the specific tasks that require coordination within a DAO and what it means to \emph{``coordinate''} in this context of blockchain require further exploration.
There is a wealth of academic literature, even if those do not specifically mention the term \emph{``DAO''} by name, including economics theory of the firm~\cite{williamson2002theory}, public choice theory~\cite{shaw2002public}, and voting paradoxes~\cite{nurmi1999voting}. They analyze the behavior of voters, interest groups, politicians, and bureaucrats in shaping policy and outcomes, which is similar to the structure of DAOs. However, voting paradoxes and the Gibbard-Satterthwaite theorem~\cite{benoit2000gibbard} demonstrate that an individual's voting power can be affected by the voting system's structure and the distribution of voter preferences, and that there is no perfect voting system~\cite{satterthwaite1975strategy} that can consistently and accurately represent the preferences of voters. Furthermore, coordination in management science is critical to ensure that resources are used efficiently and that organizations work towards the same objectives~\cite{faraj2006coordination}. Performance metrics and feedback mechanisms~\cite{faraj2006coordination}, as well as computer-based coordination tools~\cite{fish1988quilt, stokols2008ecology}, are deemed to be necessary to track progress towards organizational goals.  
Recent Literature on DAOs highlighted challenges in achieving similar goals and optimizing resources~\cite{baninemeh2023decision}, indicating for further investigation.

\subsection{
Blurred Boundary of Autonomy in DAO: Human Involvement \& External Influence
}
\label{subsec:rw-autonomous}


The concept of autonomy in DAOs has been a subject of ongoing discourse within technical, legal, and financial spheres~\cite{wright2021measuring, wright2021measuring}. The debate centers on whether DAOs should function as fully autonomous and automated entities, with no external parties involved in decision-making processes, or whether autonomy should be understood in a more nuanced way, emphasizing transparency and clear decision-making procedures~\cite{santana2022blockchain}.
Existing literature on autonomous agents, such as robots, mobile devices, as well as human-robot interactions, has identified various levels of autonomy, defined as the capacity to sense, plan, and act without human intervention or external control~\cite{sae2018taxonomy, elhannouny2019off, clough2002metrics, federal2015operation, huang2005autonomy,heckman1998liability,dowling2000intelligent}.
In the case of Decentralized Organizations, it replaces a hierarchical structure managed by humans with a protocol specified in code and enforced on the blockchain, allowing for decentralized decision-making and controlling physical property. DAOs represent a unique combination of the concepts of Autonomous Agents (without human involvement) and Decentralized Organizations (without external influence), blurring the boundaries between the two.
Specifically, DAOs are entities that operate autonomously for certain tasks using smart contracts but also rely on hiring individuals to perform specific tasks that the automaton itself cannot perform, interacting according to their protocols~\cite{ethereum}.

Defining autonomy metrics for DAOs is challenging since existing taxonomies are often ambiguous and categorical rather than quantitative along a continuum~\cite{wright2021measuring}.  Decision-making scope and liability are the most generalizable dimensions, with domain-specific definitions being less easily generalizable. Prior research on autonomous agent designs has explored ethical and liability issues, including historical liability concerns from automation to autonomous systems~\cite{heckman1998liability,dowling2000intelligent,pagallo2017automation}, and proposed varying levels of autonomy for determining ethical significance~\cite{dyrkolbotn2017classifying}.
Themes that can be applied to DAOs include the degree of human supervision, decision-making roles, liability, and ethical considerations~\cite{haupt2019artificial}. 
Ladner et al. propose an index for local autonomy in European political entities~\cite{ladner2019local}. Other dimensions, such as those related to human development considerations, may be insightful with regard to artificial intelligence discussions~\cite{ryan2019brick}, but early DAOs are not assumed to require artificial intelligence. Of these dimensions, the notions of legal autonomy and financial autonomy would seem to be applicable in the context of DAOs. This is further complicated by regulatory proposals, such as the \emph{``Responsible Financial Innovation Act,''} which classifies DAOs as business entities for tax purposes~\cite{kramerresponsible}. Additionally, the governance models of DAOs are evolving, with emerging forms such as CityDAO which proposes to adopt a mayor-council model, where an appointed Mayor holds administrative and budgetary authority, an appointed Council holds legislative authority, and Professional Managers hold administrative authority~\cite{citydao}. 
In this paper, we aim to understand the degree to which DAO operates with less human intervention and without undue external influence, potentially from corporate competitors or regulators.
\vspace{-2mm}
\section{Methodology}
\label{research_approach}
\vspace{-2mm}

To have an in-depth understanding of various DAOs, we adopt a mixed-methods approach, iteratively blending quantitative and qualitative analyses which involve multiple sources of data, including off-chain and on-chain data.

\subsection{Empirical Analysis}
We conduct an empirical analysis of DAOs at both macro and meso levels.
Our analysis begins with an exploration of on-chain and off-chain voting data, specifically measuring the frequency of new addresses and users engaging in DAOs, as well as the evolution of DAO growth, funding patterns, treasury amount, and categories of DAOs. We then focus on the governance structure of each DAO, using descriptive statistics at the macro level to identify community members involved in DAO activities such as voting and proposing, different governance mechanisms, including proposal threshold, eligibility of proposal, quorum voting, etc. 

We then use these entry points to conduct in-depth quantitative analyses, expanding upon the existing theoretical framework for investigating both on-chain and off-chain voting and proposal data.
At the meso level, we use different methods of network analysis, economic and information theory (e.g., gini, nakamoto co-efficient, entropy) presented in the literature to investigate how different factors, such as voting power, similarities and differences among DAO holders, and token holdings, are assembled to identify the positionality of DAO holders within the network, and how this influences the decentralization. We also examined the underlying smart contract to assess to which capacity DAOs operate with less human intervention and without undue external influence, potentially from corporate competitors or regulators.
We provide additional details about specific methods, and analysis results in Sections~\ref{metric-method}, \ref{section5}, and \ref{section6}. 
While we did not extensively explore the potential influence of online forum interactions on on-chain and off-chain governance decision-making in this empirical analysis, we added narratives of different soft proposal discussions where appropriate and provided suggestions for future research to combine these two information spaces in the analysis. 
We provide a summary along with the current limitations of the metrics and methods used to explore \textbf{RQ2} and \textbf{RQ3} in Appendix~\ref{metric-method}.

\subsection{Data Collection}
We collected both off-chain data including DAO proposals, voting, token, and treasury smart contracts, as well as off/on-chain voting smart contract modules. 
For DAOs having readily available source code, we turned to Etherscan~\cite{Etherscan} while it has become the de-facto source for Ethereum blockchain exploration. It offers a useful feature called \emph{``verified''} contracts, where contract writers can publish source code associated with blockchain contracts. Etherscan then independently verifies that the compiled source code produces exactly the bytecode available at a given address. 
We scraped contracts as of November 2022. 
We then checked contracts with corresponding GitHub sources to determine the correct complied version for our analysis. For MolochDAO, Meta Gamma Delta, and dxDAO, we aggregated factory contracts from DAOhaus and DAOstack.

Focusing on the network that DAOs are built upon, we searched for voting and proposal information sources on the blockchain (on-chain DAO) and the corresponding voting platform (i.e., snapshot for off-chain DAOs). If necessary, we directly access records (blocks) using a unique identifier (e.g. block ID, transaction, wallet, or contract address). As a result, we build scripts and data pipeline/techniques for data extraction in order to conduct an analysis effectively without having to sequentially walk through all the blockchain or offchain platform data. Table~\ref{tab:metadata} lists three main forms of data including smart contracts, on/off-chain proposals, and on/off-chain voting. We collected the data in a comma-separated value.

\begin{table}[]
\scriptsize
\begin{tabular}{lll}
\hline
Smart Contract Data & DAO Proposal Data                & Data Voting Data                                     \\
\hline
\begin{tabular}[c]{@{}l@{}} \\ Compound Bravo\\ DAOhaus\\ DAOstack\end{tabular} & \begin{tabular}[c]{@{}l@{}}Proposal Author\\ Proposal Title\\ Proposal description\\ Proposal Timestamp\\ Proposal start/end Block\\ Author Token Wieght\\ Proposal Signatures\\ Proposal Outcome \\ Proposal State \\ Proposal Gas Cost \\ Execution Time delay\\ intendedExecutionTime\end{tabular} & \begin{tabular}[c]{@{}l@{}}Voter Address \\ Voter Organization\\ Voter Token Weight\\ Voter Average Token Weight\\ Voter Name if available\\ Voting Pattern (1, -1, 0)\\ Voting txHash\\ Voting Gas Cost(ETH)\\ Voting Gas Cost(USDT)\\ Voter Timestamp\end{tabular} \\

\hline
\end{tabular}
\caption{List of Metadata collected for off/on chain DAOs \vspace{-2mm}}
\label{tab:metadata}
\end{table}

\vspace{-2mm}
\subsection{Formative Studies}
\vspace{-2mm}

Our formative study encompasses both individual and group sessions, comprising eight individual interviews, one group session with ten experts representing diverse DAO communities, and a break-out session with more than 10 experts in DAO governance (demographics details in Appendix~\ref{demo-formative}). The purpose of this study was to gain insights into the existing DAO governance practices. We further aim to identify and establish a set of metrics to assess the effectiveness of governance structures, as viewed through the perspective of experts, who predominantly consisted of DAO founders and leaders. 
To this end, the study sought to explore on-chain and off-chain metrics (e.g., proposal threshold, active voters) that DAO experts perceived as contributing to the evaluation of DAOs in practice and how these metrics were defined and quantified by experts.


We began by introducing ourselves and outlining our project's objective of gaining insight into how DAOs operate in practice. We then briefly presented our initial ideas on metrics for both decentralization and autonomy, such as voting patterns, proposal success, execution delay, 3rd party reliance, etc, to obtain feedback from the DAO experts. We encouraged participants to share their experiences with DAO governance and ideas for additional metrics. We also inquired about the primary goals of the specific DAOs
that they are involved with. We then asked whether there have been any significant changes to their governance policies. 
We also probed their views on the level of decentralization and autonomy in their DAOs.
Finally, we asked for their desired tools to aid in governance and the set of features they would recommend for governing and evaluating DAOs.

\vspace{-2mm}
\section{Perceptions \& Practices of DAO Experts}
\vspace{-2mm}
\label{interview}
Here, we report the result of our formative study. 

\noindent{\bf Expected level of decentralization.} Our interview findings indicate that there are varying perceptions and expectations for the level of decentralization 
across different stages of DAO development. In particular, experts suggested that early-stage DAOs often require a certain level of centralization, as \textit{``at beginning to form it, it makes a lot of sense to be more centralized and move to a less centralized over time.''}
Experts also tended to differentiate decentralization in terms of technical and social aspects, noting that--\textit{``many facets of decentralization exist (perceptual, technical, legal) and that there are tensions between abstraction and practicality of Meloc V3 contracts.''} In discussing limitations to achieving expected levels of decentralization, experts pointed to certain types of DAOs (e.g., highly concentrated towards DeFi protocol DAO) that require centralization for effective operation. Some mentioned limitations in current factory contracts and called for the exploration of better utilization of smart contracts to increase decentralization. As P1, from Blockchain association and Polychain noted-\textit{``DAOs are pretty early. Many are just copying existing protocols. Maybe decentralization could be improved with newer contract or utilizing contracts.''} 

\noindent{\bf Token-based economy lower decentralization.} We identified three metrics to evaluate decentralization, including token distribution, participation rate of token holders in voting/proposal, and members' geographies and regulatory risks.
Our experts compared the token holdings of the core team/founder to grassroots members. P3, who is involved in Aragon, Aave Hydra, DAOStar, noted that the token distribution remains largely unexplored in the context of decentralized governance-\textit{``If the founding team holds too many tokens, it disincentives others to contribute. For 'practical' decentralization, some other distribution is needed.''} 
Similarly, P2, from his experience of LexDAO, SporusDAO, CaliDAO, onchain LLC, indicated token-based governance as inefficient, hindering the achievement of true decentralization; \textit{``we're coming to the conclusion that token-based governance, is ultimately not very effective because[...] if three or four whales out of 10,000 token holders have 51\% of the power. Can you really call it decentralized?''} 
Many emphasized the importance of assessing the participation rate of token holders during voting/proposals. 
It is indicated as important to examine who is pitching the proposal and if they are from the core team. 
P6, previously worked in MakerDAO, and ConcenSys indicated that often proposals do not reach quorum without whale voters, and the proposal authors would chase down whale voters to lobby them to vote. Additionally, P8 noted that---\textit{``the reality is people don't like to participate in governance, [and] where you end up unfortunately, just speculating on the price value.''} Our experts pointed out the limitations of associating voting power with monetary value and suggested voting power to be proportional to contribution to DAOs. 

\noindent{\bf Unclear concept of autonomy.}
Our interviews revealed a recurring theme of the absence of clear metrics for autonomy within DAOs. 
While many experts view autonomy as \textit{``less human intervention, more tool use,''} others noted that different levels of autonomy depend on the functions/types of DAOs. To avoid bias and excessive human interaction, some experts suggested creating \emph{``pods''} in DAOs for different operations, such as marketing or legal. As P5, working in Polywrap DAO, Llamadao noted, \textit{``we want to create different pods and each will have different people that are governing so one pod doesn't have too much power.''}
Some experts argued that relying on human intervention is necessary due to a lack of tooling, but emphasized the need to minimize dependence on external parties. Others defined autonomy in terms of the proposal and voting process, noting that--\textit{``if proposal/voting is happening within a machine, it's autonomous.''} Finally, a few experts defined autonomy in the context of, not being controlled by the US government, noting that---\textit{``I don't think it should honestly have to do with if people or machines can do quick execution. You can mean prolifically autonomous, is it not controlled by, say, US government.''} 

\noindent{\bf Limitations of current tooling to enable autonomy.} Experts did not mention many solutions but instead suggested limitations in building autonomy in DAOs. Some of the limiting factors include execution time delays of DAO operations due to a lack of tooling, and proposals going into backlog due to quorum votes. P8, from DAOHQ noted that companies are attempting to improve user experience on the interface to make it easier for governance facilitators to support the process in a more automated, but there are not many options available. P7, involved in pleasr DAO, NounsDAO added that limitations around token holding and lack of interest in participating in governance often cause delays in the proposal process, since there are specific vote requirements to move the proposal forward. In this word---\textit{``Token distribution among a few or core members means a single point of failure, causing proposal delay.''}

\noindent{\bf Contextual factors making whales to vote in similar way.}
We identified several factors that might have contributed to similar voting patterns among whale voters. Some interviewees noted that the majority of whales belong to the managerial team leading to vote in a similar way. 
P5 noted---\textit{``so these whale,
they started with a managerial, centralized team and so they are the largest token holders.''} Additionally, P5 added---\textit{``some of their communication happen through private channels like private discord, DMs or telegrams.''}, which may contribute to the similar voting patterns observed.  
Another factor contributing to this trend is people's trust in the voting of earlier whales. P8 noted that---\textit{``Honestly, for many proposals when earlier whales vote in a certain way, a lot of people just follow [..]many don't care to read through proposal. But they trust, whales know what they're doing.''} Peer pressure during the voting process was also mentioned as a factor that may lead individuals to vote in a similar way to the whales. 
P8 noted---\textit{``I didn't want to vote against the proposal, Just because, my reputation would be on the line or something.''} Overall, 
\emph{whales running governance} is identified as a major limitation in DAOs.

\noindent{\bf Need for qualitative investigation alongside quantitative evidence.} Participants suggested that contextual factors should be incorporated including community sentiment from social channels, timing and the situation in which people join DAOs, additional financial metrics from existing exchanges, and the type of proposals being considered. 
While higher proposal thresholds 
pointed out the trade-offs of protecting against spam or malicious proposals, 
P9, from Gitcoin highlighted the practical implications of the threshold on the impact on DAO token value in terms of market capitalization.
\vspace{-2mm}\section{Summary of 10 DAOs}
\label{section4}
\vspace{-2mm}

In this paper, we selected 10 DAOs, which were formed between 2019 and 2022, prioritizing diverse categories (e.g., investment, defi, video, social, etc), popularity, and market capitalization. To add diversity for governance design, we, in particular, considered four DAOs, created with popular DAO factory contracts, including Governance Bravo 
DAOhaus, Kleros, and DAOStack, in the governance structure.
Table~\ref{tab1:macro} presents the key parameters of the 10 DAO projects, including the treasury size, number of DAO holders, active voters, total votes, participation rate in proposals, number of proposals, number of proposal creators, quorum voting, and proposal threshold (more details in Appendix~\ref{10_dao}). 

 \begin{table*}[]
 \scriptsize
\begin{tabular}{llllllllllll}
\hline
DAO               & Category                                               & Treasury & \#Holders & \#Active & \begin{tabular}[c]{@{}l@{}}Avg \\ votes\end{tabular} & Participation \% & \# proposal & \#votes & \#Creator & \begin{tabular}[c]{@{}l@{}}Proposal \\ Threshold\end{tabular} & Quorum\\
\hline
CompoundDAO       & Protocol                                               & 90.7M    & 205.9k    & 4.1k     & 78                                                           & 1.80\%           & 137         & 12750   & 48        & \begin{tabular}[c]{@{}l@{}}25000\\ COMP\end{tabular}          & 400,000                                                \\

BitDAO            & Investment                                             & 1.6B     & 20.4k     & 320      & 53                                                           & 1.30\%           & 19          & 1347    & 14        & \begin{tabular}[c]{@{}l@{}}200,000 \\ BIT\end{tabular}        & 100,000,000                                                                            \\
AssangeDAO        & Social                                                 & 242.9k   & 6.3k      & 1.1k     & 298                                                          & 12.30\%          & 11          & 3576    & 2         & \begin{tabular}[c]{@{}l@{}}1,000\\  JUSTICE\end{tabular}      & 867,576,97                                                                             \\
Proof of Humanity & Social                                                 & 801.7k   & 35.2k     & 3.1k     & 153                                                          & 8.80\%           & 104         & 16557   & 25        & 0 VOTE                                                        & NA                                                                                     \\
Bankless DAO      & Web3 fund                                              & 1.6M     & 5.9k      & 3,375    & 344                                                          & 43\%             & 51          & 18272   & 7         & 0 BANK                                                        & NA                                                                                    \\
KrauseHouse       & Sports                                                 & 1.2M     & 1.8k      & 595      & 32                                                           & 30\%             & 131         & 4,484   & 5         & \begin{tabular}[c]{@{}l@{}}1,000 \\ KRAUSE\end{tabular}       & 0                                                                                      \\
LivePeer          & Video                                                  & 0        &   NA        &    NA      &           NA                                                   &     NA             & 7           & 600     & 7         & 0 LPT                                                         & 1/3*27,004,976                                                                         \\
MetaGammaDelta    & Social                                                 & 29.7k    &  NA         & 49       & 2                                                            & 29.40\%          & 79          & 132     & 30        & 1 DAI                                                         & 0                                                                                      \\
MolochDAO         & \begin{tabular}[c]{@{}l@{}}Public \\ Good\end{tabular} & 0.1      & NA          & 60       & 4                                                            & 50.80\%          & 32          & 83      & 29        & 1 WETH                                                        & 0                                                                                      \\
dxDAO             & Protocol                                               & 81M      & 1.4k      & 232      & 2                                                            & 5.80\%           & 864         & 3161    & 160       & 0 rep                                                         & \begin{tabular}[c]{@{}l@{}}0 if boosted \\ else 50 \% REP\\  token supply\end{tabular}\\
\hline
\end{tabular}
\caption{DAO Projects Summary in terms of macro information\vspace{-2mm}}
\label{tab1:macro}
\end{table*}

Among 10 DAO projects, five perform off-chain voting (BitDAO, AssangeDAO, Bankless, KrauseHouse, and Proof of Humanity), while the rest perform on-chain. Gnosis Safe~\cite{gnosis} Snapshot platform is often used for off-chain voting aggregation for its user-friendly interface. However, it is not implemented in blockchain and requires further action from the operation team through multi-sig~\cite{daohaus}.
Across 10 different DAOs, the requirements vary significantly. For instance, BitDAO and CompoundDAO only allow whitelisted addresses to propose, and proposal submissions require a specific amount of tokens above a certain threshold. However, the threshold is often too high for grassroots users\footnote{The ``grassroot users'' cluster represents addresses with token holding that is less than the whale address token threshold.} to propose without delegation from others, resulting in industry entities or large token holders submitting most proposals. AssangeDAO, BanklessDAO, and KrauseHouse only permit authors verified on snapshot to propose. 
In Livepeer, anyone with a sufficient amount of staked\footnote{Staked governance tokens are locked for a period of time. Those that stake their tokens often bear certain protocol risks.} governance tokens can propose, while in Meta Gamma Delta and MolochDAO, proposals require sponsorship\footnote{Sponsoring proposals in Moloch and Meta Gamma Delta DAO is an act in which a DAO shareholder moves the proposal into the voting pipeline.} from a member. Lastly, dxDAO has the most flexible proposal requirement, allowing anyone to propose. Each DAO also has a specific quorum voting strategy for accepting proposals, with CompoundDAO requiring 400,000 COMP, BitDAO requiring 100,000,000 BIT, AssangeDAO requiring 867,576,975 JUSTICE tokens, etc.

\vspace{-2mm}
\section{RQ2: Decentralization of DAOs}
\label{section5}
\vspace{-2mm}

In this section, we investigated the level of decentralization in the current DAOs with respect to voting and proposal.

\vspace{-2mm}
\subsection{Voting Patterns Among DAO Holders}
\vspace{-2mm}

Our interview study (section~\ref{interview}) highlights the limitation that token-based governance disincentivizes grassroot holders from participating in the decision-making process, particularly with regard to the tendency for core or managerial teams to hold a majority of tokens, leading to centralized decision-making by a small group. 
Given this, we examine the interaction between different DAO holders and present clustering of holders based on their voting decisions for different proposals. 
We also analyze the voting patterns of large token holders (whales) to establish their positionalities and investigate whether their voting patterns align with broader community. Given that greater token holdings represent greater influence of holders in DAO voting systems, if voting patterns of whale addresses do not align with grassroots users or their voting patterns only have similarities among themselves, it implies that the grassroots users are marginalized, suggesting poor decentralization.

\begin{figure*}[!htb]
\begin{subfigure}{0.20\textwidth}
  \includegraphics[width=\linewidth]{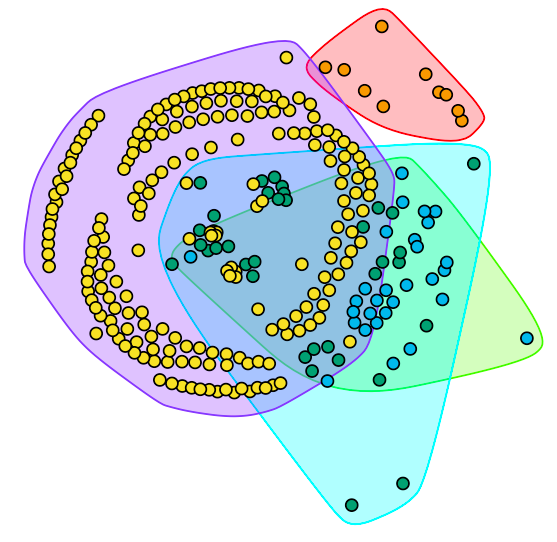}
  \caption{BitDAO}\label{fig:bit_image1}
\end{subfigure}\hfill
\begin{subfigure}{0.20\textwidth}
  \includegraphics[width=\linewidth]{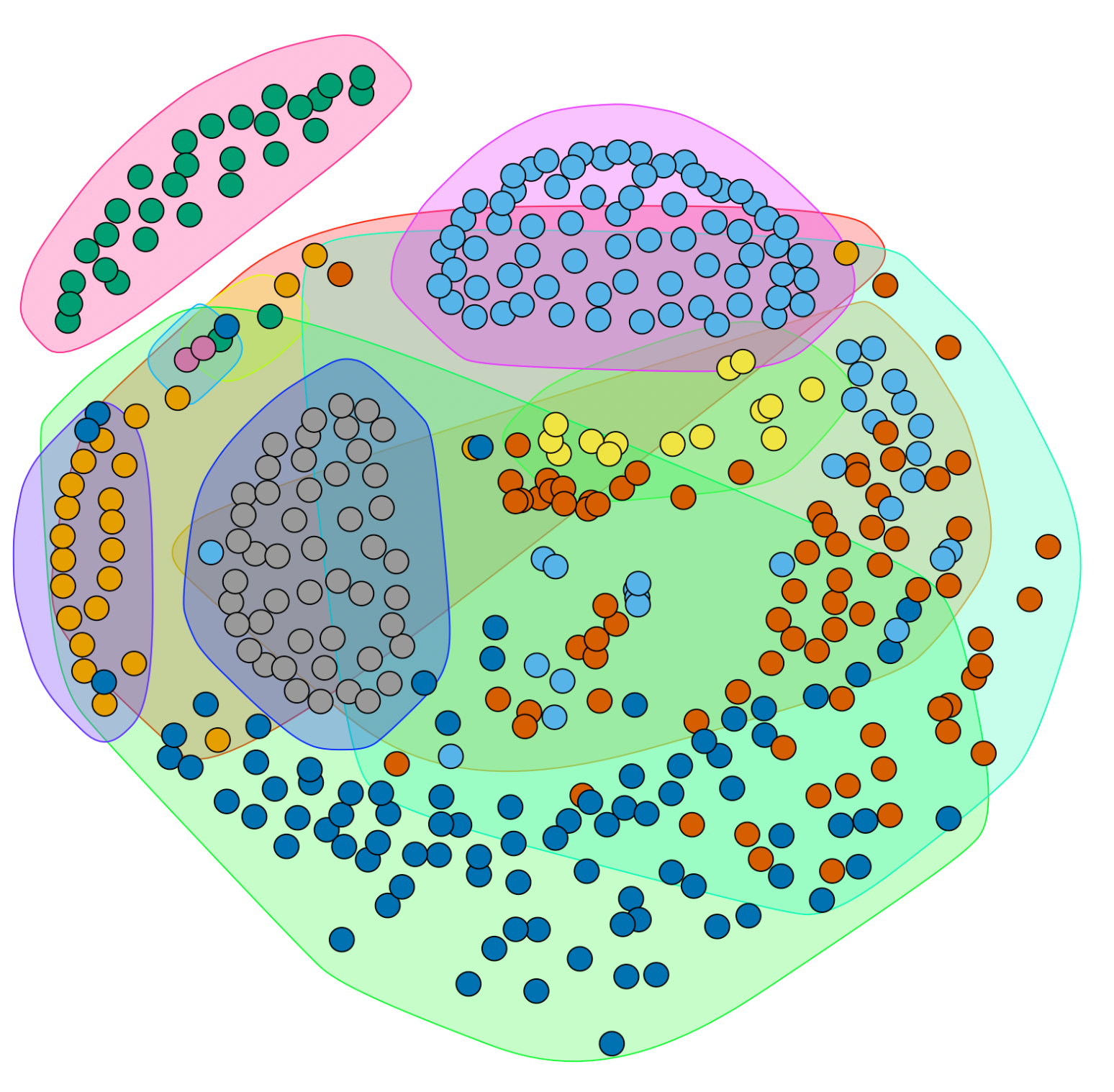}
  \caption{CompoundDAO}\label{fig:comp_image2}
\end{subfigure}\hfill
\begin{subfigure}{0.20\textwidth}%
  \includegraphics[width=\linewidth]{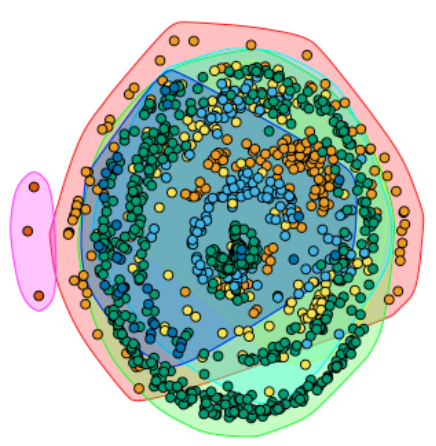}
  \caption{AssangeDAO}\label{fig:assange_image3}
  \end{subfigure}\hfill
\begin{subfigure}{0.20\textwidth}%
  \includegraphics[width=\linewidth]{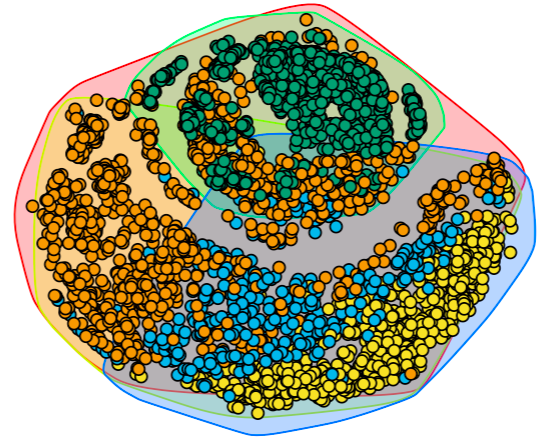}
  \caption{BanklessDAO}\label{fig:assange_image3}
  \end{subfigure}\hfill
\begin{subfigure}{0.20\textwidth}%
  \includegraphics[width=\linewidth]{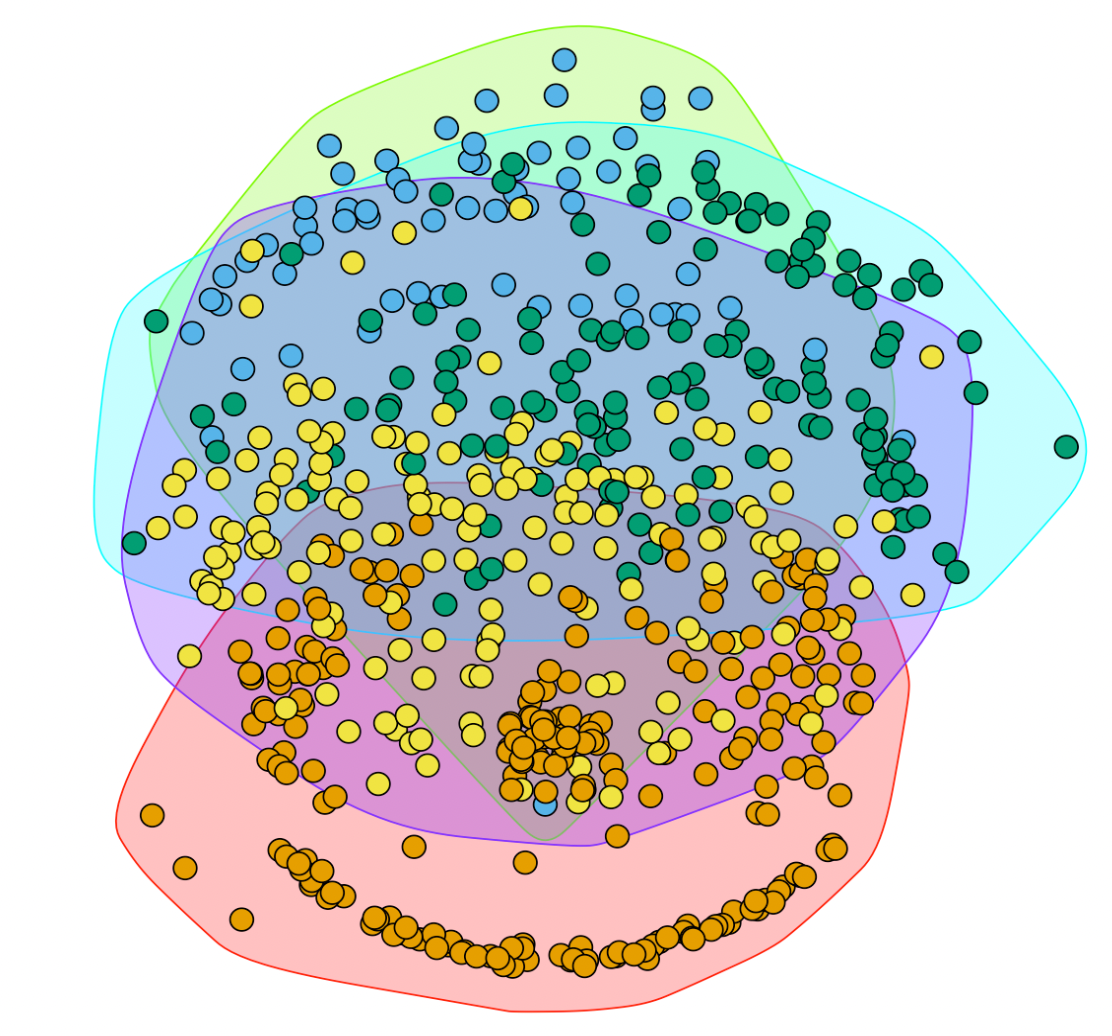}
  \caption{KrauseHouse}\label{fig:assange_image3}
\end{subfigure}
\caption{DAO holders' voting behavior for different proposals are displayed as community graphs. A cluster is defined based on their voting behavior where 1: voted for certain proposal, -1: voted against certain proposal. The colored polygons in subgraphs (a), (b), and (c) represent the nodes with similar voting behavior. There are some overlapped colored regions which indicate that nodes in those clusters have some voting similarity with nodes in other cluster.
For example, figure (a) shows no overlapped clusters because the nodes from different clusters do not share any similar voting behavior. In contrast, figure (b) shows the cluster containing yellow nodes has a overlapped region with the cluster containing red nodes due to some voting similarity (at least one voting similarity between nodes/addresses). 
\vspace{-2mm}} 
\label{cluster1}
\end{figure*}

\begin{figure*}[!htb]
\begin{subfigure}{0.25\textwidth}
  \includegraphics[width=\linewidth]{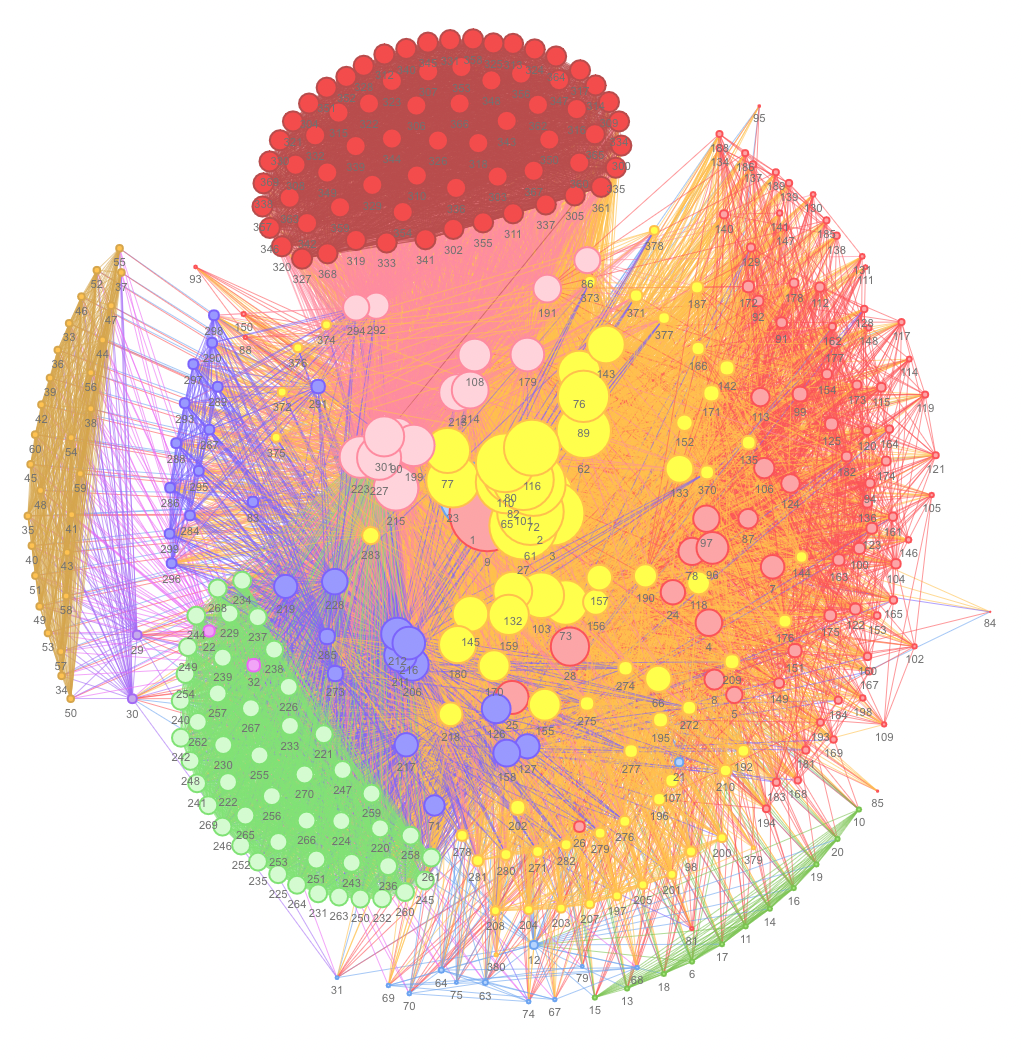}
  \caption{CompoundDAO Voting Pattern with directed graph. Each edge means the two nodes voted the same way to a proposal.}\label{fig:c-1}
\end{subfigure}\hfill
\begin{subfigure}{0.25\textwidth}
  \includegraphics[width=\linewidth]{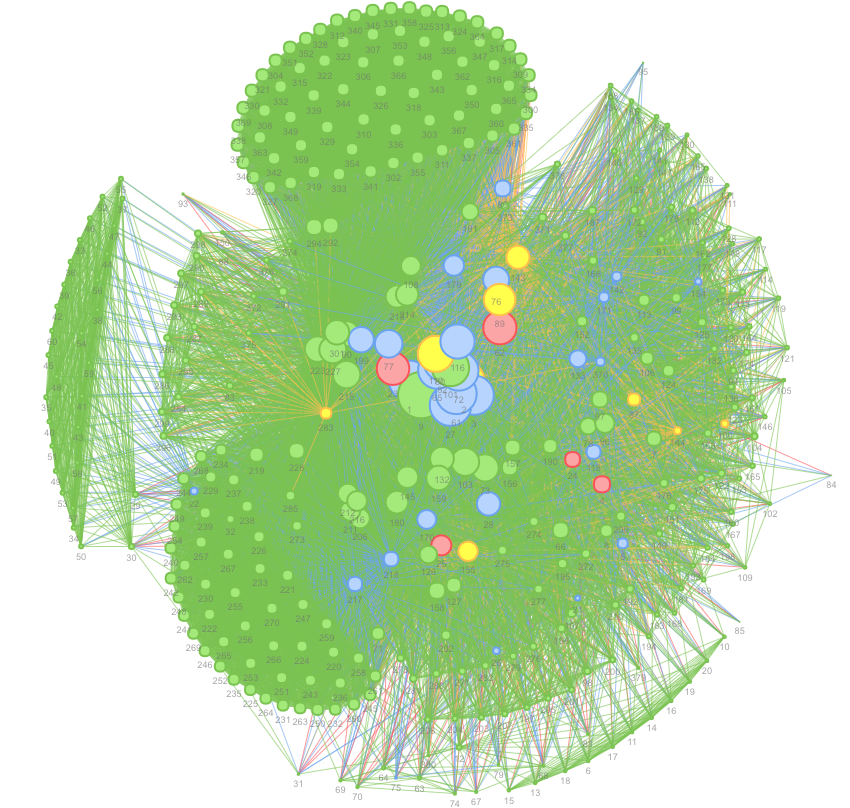}
  \caption{CompoundDAO Voting Pattern of different type of users. Blue represents Industry. Yellow: non profit. Red: Educational Institute. Green: individual users, big green nodes are KOLs. 
  }
  \label{fig:c-2}
\end{subfigure}\hfill
\begin{subfigure}{0.25\textwidth}%
  \includegraphics[width=\linewidth]{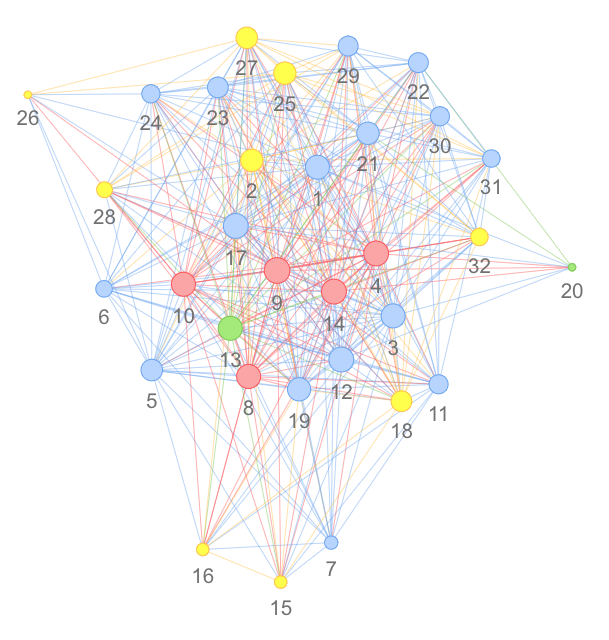}
  \caption{CompoundDAO Whale address. Industry:blue, Non Profit:green, Educational Institute:red, Key Opinion Leader:yellow }\label{fig:c-3}
\end{subfigure}
\caption{
We present CompoundDAO as an example to demonstrate the result of the first hypothesis, H1. Here the size of the nodes is directly proportional to their input degree (i.e., the larger a node, the more tokens it has accumulated in the DAO). Blue nodes represent industry,  green for non-profit organizations, red for educational institutes, and yellow for key opinion leaders (KOLs). 
We found 32 addresses with a large threshold of Compound token holding. 
Figure 2c shows that whale addresses from industry, KOLs and educational institutes tend to have similar voting behavior.\vspace{-2mm}}  
\label{cluster-2}
\end{figure*}

\begin{table*}[]
\scriptsize

\begin{tabular}{lllllllllll}
\hline
DAOs        & \begin{tabular}[c]{@{}l@{}}Freq of \\ community\end{tabular} & Vertics & Edge    & Diameter & \begin{tabular}[c]{@{}l@{}}Mean\\ Distance\end{tabular} & 
\begin{tabular}[c]{@{}l@{}}Degree \\ Transitivity\end{tabular} & \begin{tabular}[c]{@{}l@{}}Degree \\ Assortativity\end{tabular} & \begin{tabular}[c]{@{}l@{}}Degree \\ Centralization\end{tabular} & \begin{tabular}[c]{@{}l@{}}Betweenness \\ Centralization\end{tabular} & \begin{tabular}[c]{@{}l@{}}Closeness \\ Centralization\end{tabular} \\
\hline
Bit         & 4                                                            & 283     & 26985   & 4        & 1.33                                                                                                                & 0.93                                                           & 0.616                                                            & 0.31                                                             & 0.04                                                                  & 0.44                                                                \\
Assange     & 6                                                            & 1065    & 380410  & 3        & 1.33                                                                                                                & 0.87                                                           & 0.04                                                           & 0.32                                                             & 0.002                                                                 & 0.45                                                                \\
PoH         & 5                                                            & 3084    & 1403628 & 4        & 1.7                                                                                                              & 0.66                                                           & -0.2                                                           & 0.68                                                             & 0.008                                                                 & 0.77                                                                \\
Bankless    & 5                                                            & 3139    & 1758001 & 3        & 1.64                                                                                                              & 0.59                                                           & 0.09                                                           & 0.64                                                             & 0.004                                                                 & 0.24                                                                \\
KrauseHouse & 4                                                            & 587     & 63713   & 3        & 1.63                                                                                                              & 0.74                                                           & 0.07                                                           & 0.54                                                             & 0.019                                                                 & 0.59                                                                \\
Compound    & 10                                                           & 2482    & 378635  & 4        & 1.65                                                                                                              & 0.75                                                          & 0.026                                                          & 0.75                                                             & 0.09                                                                  & 0.72                                                                \\
LivePeer    & 2                                                            & 367     & 21802   & 2        & 1.68                                                                                                             & 0.68                                                           & -0.2                                                           & 0.67                                                             & 0.04                                                                  & 0.79                                                                \\
MGD         & 2                                                            & 15      & 56      & 5        & 1.49                                                                                                                & 0.66                                                           & -0.2                                                            & 0.32                                                             & 0.09                                                                  & 0.42                                                                \\
Moloch      & 2                                                            & 14      & 50      & 3        & 1.45                                                                                                              & 0.65                                                           & -0.29                                                           & 0.37                                                             & 0.13                                                                  & 0.5                                                                 \\
dxDAO       & 4                                                            & 270     & 35520   & 2        & 1.02                                                                                                           & 0.99                                                           & 0.60                                                           & 0.32                                                             & 0.01                                                                  & 0.03          \\                                                     
\hline
\end{tabular}
\caption{Network Graph connectivity and clustering statistics for each DAOs in terms of reciprocity, transitivity and assortativity, vertices, edges, diameter, mean distance}
\label{tab:graph}
\end{table*}

\vspace{-2mm}
\subsubsection{Graph Model}
We developed network graphs to understand DAO holders' voting patterns across different proposals, considering their voting weights. 
In the graphs, each node $v_i (\in V)$ represents a unique Ethereum address, where $V$ indicates the full set of nodes. Nodes are connected via an edge if they voted the same way (e.g., voted yes) in the same proposal, and $E$ denotes a set of edges $\{(v_{i}, v_{j}) | v_{i}, v_{j}\in V\}$, where $(v_{i}, v_{j})$ indicates an edge between $v_i$ and $v_j.$
The graph structures of DAOs (Figures~\ref{cluster1} and \ref{cluster-2}) are derived using the walktrap algorithm, commonly used to identify communities in networks via random walks~\cite{petrochilos2013using}. 
Here, the random walks are also used to compute distances between nodes. Therefore, we calculated the optimal similarities and differences of voting behaviors of DAO holders in a weighted graph $G(V,E)$.

The colors represent clusters related to \textit{``VoterType''} who vote in similar ways for different proposals\footnote{We use the igraph R package to calculate the different clusters: https://igraph.org/r/}. 
%
If two nodes have more than one same vote, they are more likely to be in the same cluster. 
%
To construct a voting network graph, we first created a matrix\footnote{We used an incidence matrix to make a bipartie graph since the number of rows is different than the number of columns. Within the $incidence matrix$, there will be an adjacency matrix with similar number of columns and rows.} whose entity indicates a vote on a certain proposal done by an address.
Entity values 1 and -1 imply a vote for and against a certain proposal, respectively. 
Based on the matrix, we built an adjacency matrix where columns and rows represent ``Votes'' and ``Addresses'', respectively. Then we constructed the bipartite graph between ``Address'' and ``Votes'' from the adjacency matrix. In the graph, an edge between Addresses and Votes indicates that an address has voted either against or for the proposals. 

Further, we present the community cluster with voting weight $w(v_{i}, v_{j})$, 
%
which corresponds to the token amount each address delegated for themselves or others to perform the transaction (i.e., to vote for the proposal). 
We calculated the weight of each address by averaging the voting weight (or token amount) delegated to the address for each proposal. Nodes are sized by the number of votes/tokens they delegated; that is, larger nodes represent more highly delegated token accounts (Figure~\ref{cluster-2}). 



\vspace{-2mm}
\subsubsection{Patterns of Voting Networks} 

We present the main static properties of the constructed graphs. Table~\ref{tab:graph} summarizes the network properties of 10 DAOs: number of nodes and edges, diameter, mean distance, degree distributions, density, component clustering coefficients, and assortativity. 
Note that here we considered only addresses with an edge by eliminating null votes made by different addresses (i.e., discarding addresses that have not voted on any proposal). 
%

In Figure~\ref{cluster1}, the clusters are colored based on the optimal similarity of voting behaviors and the distance between the nodes. The overlapped regions of colored polygons indicate that there is at least one voting similarity between the nodes in one cluster and the others. In BitDAO, nodes in specified clusters have very few similarity in voting with other clusters, which results in comparatively less overlapping regions in Figure~\ref{fig:bit_image1}. On the other hand, the CompoundDAO network (Figure~\ref{fig:comp_image2}) has a total of 10 clusters, where some nodes in clusters have at least one similar voting with nodes in other clusters. For AssangeDAO (Figure~\ref{fig:assange_image3}), there are a total of six clusters, and most nodes have at least one voting similarity with other clusters. 
As such, one can see the significant difference in the network structures between DAO, which implies heterogeneous community voting behaviors. 
In contrast, we observed less cluster diversity in proof of humanity where the number of voters are second highest, however, voters tend to have similar voting patterns 
(Figure~\ref{cluster-new} in Appendix). Meta Gamma Delta, Moloch, Livepeer, dxDAO tend to follow the similar trend.

\noindent\textbf{Connectivity and clustering properties.}
The assortativity coefficient~\cite{meghanathan2016assortativity} measures the homophyly level of the graph (i.e., how nodes are connected for a given property), and it ranges in $[-1,1]$. More specifically, according to \cite{meghanathan2016assortativity}, a graph is said to be strongly assortative, weakly assortative, neutral, weakly disassortative, and strongly disassortative, if the assortativity coefficient falls into the ranges $[0.6,1]$, $[0.2,0.6)$,$(-0.2,0.2)$,$(-0.6,-0.2]$, and $[-1,-0.6]$, respectively. 
Table~\ref{tab:graph} shows that majority of the DAO, including CompoundDAO, Krausehouse, Bankless and Assange, PoH to be neutrally assortative while BitDAO and dxDAO that are strongly assortative indicating that the nodes tend to be connected to other nodes with similar properties. However, in case of neutral assortativity, 
many DAO do not have the trend that members vote the same way as others with similar characteristics to them. We further illustrate this property in conjuction with degree centrality where we observed PoH, Bankless, Compound, Livepeer having degree centralization value above 0.6 indicating a significant concentration of node degrees. None of the DAO has degree centralization value below 0.3. 
The transitivity coefficient (also known as a clustering coefficient) measures the probability for nodes connected via multiple edges to be adjacent; in other words, if there are two links $(v_{i},v_{j})$ and $(v_{j},v_{k})$, what the probability of there being a link $(v_{i},v_{k})$ is. Table~\ref{tab:graph} illustrates various levels of transitivity across DAOs and the value is quite high. dxDAO has the highest transitivity at 0.99, implying that similarity in voting behaviors between nodes connected to other addresses is 99\%, BitDAO which in 93\%, AssangeDAO(87\%), CompoundDAO(75\%), KrauseHouse(74\%). PoH, Livepeer, MGD and Moloch have the probability for nodes connected via multiple edges 65-68\%. 

We further provide a detailed examination of the voting network characteristics of various DAOs. Specifically, our analysis of the CompoundDAO protocol revealed high similarities in voting behaviors among grassroot users. Additionally, we identified one cluster composed of nodes with higher token weight (referred to as \emph{whale addresses}) that exhibit similar voting patterns. These nodes will be discussed in further detail in Section~\ref{h1}. Our analysis of the BitDAO also revealed similar voting patterns among nodes with higher token holdings, with less cluster distance and a tendency to vote in a similar manner. However, some higher token holders in these DAOs tend to be representative of grassroots compared to those in the CompoundDAO, although they remain polarized \footnote{Polarized voting refers to the tendency for dao voters to align themselves with clusters while token holdings dictate this polarization 
}. We also observed significant overlaps in voting behaviors among different token holders in dxDAO, with only a few proposals resulting in significant voting leading to two small clusters. Similar patterns were observed in the AssangeDAO and BanklessDAO, with polarized voting behaviors among token holders. 



In contrast, our analysis revealed a more equitable distribution of voting patterns among different types of token holders for the Proof of Humanity DAO.
It might be because majority of the token holders ($>99\%$) in this DAO are grassroot users.
For Livepeer, we found two main different types of voting behaviors with small overlaps regardless of the token holdings of voters, implying better decentralization and less biased voting patterns dictating upon token holding status. Finally, Meta Gamma Delta (MGD) and MolochDAO, we can not say that their decentralization level is high due to the small number of voters, however, token holding didn't dictate the voting patterns. 



\begin{tcolorbox}[width=\linewidth, colback=white!95!black, boxrule=0.5pt, left=2pt,right=2pt,top=1pt,bottom=1pt]
\stepcounter{observation}
{\bf Takeaways \arabic{observation}:}
{CompoundDAO, AssangeDAO, Bankless, and Krausehouse protocols exhibit high polarization based on token holdings while
Proof of Humanity, MGD, and Moloch protocols exhibit more equitable distribution.

}

\end{tcolorbox}


\subsubsection{Positionality of Whale Addresses}
\label{h1}

We identify the whale addresses in different DAO by using Complementary Cumulative Distribution Function (CCDF: the likelihood that the random variable is above a particular level, defined as $1-CDF$), based on DAO holders' token holdings. 
As an example, we present CompoundDAO's member addresses for the network analysis. Table~\ref{tab:graph-whale} in Appendix present number of whale for 10 DAOs and other network properties. We find that there are $2482$ unique addresses in CompoundDAO, from which $32$ addresses (Figure~\ref{fig:c-3} are identified as whale addresses ($100k-10m COMP$). We analyze node types and voting behaviors, which allows us to identify the different characteristics between whales and grassroots.
Figure~\ref{fig:c-2} shows $10$ clusters of CompoundDAO nodes, calculated based on their voting behavior and weight. 
Moreover, considering the information obtained from the CompoundDAO off-chain governance voting profile, we identify the ``type of organizations''. 
We color the nodes depending on their types, where green represents ``grassroot users,'' blue is ``industry,'' red is ``educational institute,'' and yellow is ``non-profit''. 
From Figure~\ref{fig:c-2}, one can see that non-profit organizations tend to agree with the grassroot and broader communities regarding votes. The blue nodes ``industry'' tend to vote in a manner that are less likely to be people representative or have different voting patterns from ``grassroot.'' ``Education Institute'' seems to have similar voting patterns to ``industry.'' 

In Figure~\ref{fig:c-2}, 90.6\% of the green nodes have less than 0.25 tokens. From green nodes, we found a few users (don't represent any industry entity) having tokens in the whale threshold and have a certain influence in the CompoundDAO community (inferred from discourse conversation~\cite{comp}
Thus, we name them as ``\textbf{Key opinion leader} \footnote{A key opinion leader (KOL) is a trusted, well-respected influencer with proven experience and expertise, here in DAO community.}'' and presented those nodes with bigger thresholds of token weights.
Figure~\ref{fig:c-3} shows graph connectivity (voting similarity) among nodes with significant delegated voting weight that is, whale addresses that fall into one of the following categories, ``Key Opinion Leader,'' ``Industry,'' ``Non-Profit,'' and ``Educational Institute''. 
These addresses significantly contribute to the CompoundDAO proposal's success (i.e., accept) or failure (i.e., reject). 
%
The nodes are connected to each other via a different colored edge, depending on their similarity in voting behaviors. 
%
An address/node can represent a person or a group of people (of different autonomous organizations). However, due to the lack of user profiles (unlike CompoundDAO), it is hardly possible to differentiate the category of organization in many other DAOs we explored.

\subsection{DAO Holders' Proposal Pattern}
\label{proposal}
In this section, we analysed of DAO holders' proposal data to investigate the potential influence of various factors on proposal outcomes. 

\subsubsection{Factors Influencing Proposal Outcome} 
\label{proposal-find}
We first performed logistic regression of the proposal outcome, where the predictor is the amount of token holdings of proposers at the time of the proposal. We statistically confirm if the amount of token holdings dictates the outcome of the proposal. 
Second, we investigated the potential relationship between token holdings and proposal success. To this end, a mixed-effect regression analysis was conducted on the proposal outcome, with token amount as a fixed effect and id (address) as a mixed effect variable. The significance of this model was then evaluated using an ANOVA test, in comparison to a base model that only included id as a mixed effect and did not consider token amount as a fixed effect. This test was conducted under the assumption that \emph{token holdings may increase the possibility of success.} Additionally, we run another significance test for proposal outcome, comparing it to a base model that did not include id as a mixed effect, in order to investigate the potential influence of author address on proposal success. 





We find a significant positive correlation between the \textit{\textbf{``token amount''}} and the proposal outcome (p-value<0.05) in CompoundDAO, dxDAO and krauseHouse,. In particular, for CompoundDAO, we find that token holdings have a significant effect ($p-value = 0.00676$) with an odd ratio of $95\% CI [1.244048, 1.000015]$. Further, the mix-effect analysis also confirms the token holding's significance to proposal success with the ANOVA test ($Pr(>X^{2}): >0.05$). 
For contextual information, total 128 proposals submitted to CompoundDAO between May 1st, 2020 and December 5th, 2022 revealed that industry leaders and key opinion leaders (KOLs) were responsible for 101 of these proposals. Of these, 90\% were submitted by authors who held significant amounts of tokens. Overall, proposals submitted by organizations (e.g., Alameda Research)
 and KOLs (e.g., Arr0)
had an aggregate success rate of 88.9\%. In contrast, proposals submitted by "grassroot users" had a success rate of 50\%.

Similarly, dxDAO (p-value: $7.25e^{-07}$) and KrauseHouse (p-value: $0.00054$) also show the significance of token holdings to the proposal's success. The mixed effect analysis with the assumption of token holding to increase the proposal success possibility also confirms the significance for dxDAO ($Pr(>X^{2}): 0.01247$). Further, the mixed effect analysis with the assumption of author address to increase the proposal success possibility also shows the significance for dxDAO ($Pr(>X^{2}): 2.2e^{-16}$). On the other hand, we did not observe any significance of token holdings for BitDAO, BanklessDAO, and AssangeDAO.
However, there could be other contextual factors; for example, in BitDAO, there are predefined whitelisted addresses, and addresses that own more BIT than the proposal threshold (200000 BIT) can only propose.
Therefore, homogeneity across token holdings can limit model functionality. Furthermore, BitDAO shows a significant result ($Pr(>X^{2}): 2.2e^{-16}$) of \textit{\textbf{``author address''}} as a mixed effect to increase the possibility of proposal success. Similarly, for BanklessDAO, there are predefined authors selected to propose, while only the addresses from the consensus layer in AssangeDAO can propose. Proof of Humanity, Meta Gamma Delta, and MolochDAO where anyone can propose, showed no significance for \textit{``token holding''} with proposal success. 
Although people can delegate their votes, the total vote is $1+\sqrt{delegated vote}$ for PoH having equal number of token to vote. For example, 16 addresses delegate to an address, then the total vote for that address would be $1+4=5$. This means there is some level of homogeneity across token holding. However, for Proof of Humanity ($Pr(>X^{2}): 2.2e^{-16}$) and Meta Gamma Delta ($Pr(>X^{2}): 1.02e^{-10}$), we found a significance of \textit{``author address''} to increase the possibility of proposal success.

  
\begin{tcolorbox}[width=\linewidth, colback=white!95!black, boxrule=0.5pt, left=2pt,right=2pt,top=1pt,bottom=1pt]
\stepcounter{observation}
{\bf Takeaways \arabic{observation}:}
{Proposal success has a significant positive correlation with token holdings by proposers for KrauseHouse, CompoundDAO, and dxDAO. However, BitDAO, BanklessDAO, and AssangeDAO showed no significant correlation due to predefined whitelisted authors.
}

\end{tcolorbox}

\subsection{DAO Holders’ Voting Power}

In this section, we estimate the decentralization level regarding holders' voting power.

\subsubsection{Evaluation of Decentralization Levels}

First, we conducted a correlation analysis between voters' token weight and a participation rate, which is defined as the number of proposals that a voter participated in. We aim to determine whether the participation rate, therefore the voting power, is biased toward rich token holders. Here, the positive correlation between the participation rate and token weight may imply worse decentralization due to the more active participation of only richer voters than the others. Table~\ref{tab:decen-level} summarizes a Pearson correlation coefficient and p-value of 10 DAOs. While the results were mixed, five systems - CompoundDAO, BitDAO, AssangeDAO, Proof of Humanity, and LivePeer - showing a significant positive correlation. MetaGammaDelta, on the other hand, displayed a significantly strong negative correlation. Consequently, it appears that in many DAOs, wealthy voters are more likely to participate, exacerbating poor decentralization.


\begin{table*}[]
\small
\begin{tabular}{lllllllll}
\hline
Name              & Corr. & P-value   & \begin{tabular}[c]{@{}l@{}}Entropy \\ (Token)\end{tabular} & \begin{tabular}[c]{@{}l@{}}Entropy\\ (Voting score)\end{tabular} & \begin{tabular}[c]{@{}l@{}}Gini\\ (Token)\end{tabular} & \begin{tabular}[c]{@{}l@{}}Gini\\ (Voting score)\end{tabular} & Nakamoto & \begin{tabular}[c]{@{}l@{}}The num. \\ of voters\end{tabular} \\
\hline
CompoundDAO       & 0.27  & 3.32E-44  & 5.86                                                       & 5.17                                                             & 0.98                                                   & 0.99                                                          & 15       & 2483                                                          \\
BitDAO            & 0.21  & 0.0005    & 3.9                                                        & 3.35                                                             & 0.96                                                   & 0.97                                                          & 6        & 284                                                           \\
AssangeDAO        & 0.24  & 4.69E-15  & 7.83                                                       & 7.42                                                             & 0.84                                                   & 0.87                                                          & 50       & 1066                                                          \\
Proof of Humanity & 0.38  & 2.18E-108 & 11.46                                                      & 8.85                                                             & 0.05                                                   & 0.73                                                          & 1448     & 3086                                                          \\
Bankless DAO      & 0.05  & 0.003     & 9.95                                                       & 9.13                                                             & 0.7                                                    & 0.84                                                          & 268      & 3234                                                          \\
KrauseHouse       & 0.05  & 0.15      & 5.59                                                       & 4.68                                                             & 0.91                                                   & 0.95                                                          & 7        & 592                                                           \\
LivePeer          & 0.5   & 3.57E-27  & 5.36                                                       & 4.63                                                             & 0.91                                                   & 0.95                                                          & 9        & 368                                                           \\
MetaGammaDelta    & -0.89 & 1.99E-07  & 4.22                                                       & 3.62                                                             & 0.03                                                   & 0.51                                                          & 9        & 19                                                            \\
MolochDAO         & 0.18  & 0.51      & 2.37                                                       & 1.823                                                            & 0.73                                                   & 0.81                                                          & 2        & 15                                                            \\
dxDAO             & 0.03  & 0.57      & 6.21E-12                                                   & 1.25E-11                                                         & 0.99                                                   & 0.99                                                          & 1        & 278     \\
\hline
\end{tabular}
\caption{Evaluation of decentralization levels through correlation analysis and three metrics: Entropy, Gini coefficient, and Nakamoto.}
\label{tab:decen-level}
\end{table*}

\begin{figure*}[!htb]
\begin{subfigure}{0.45\textwidth}
  \includegraphics[width=\linewidth]{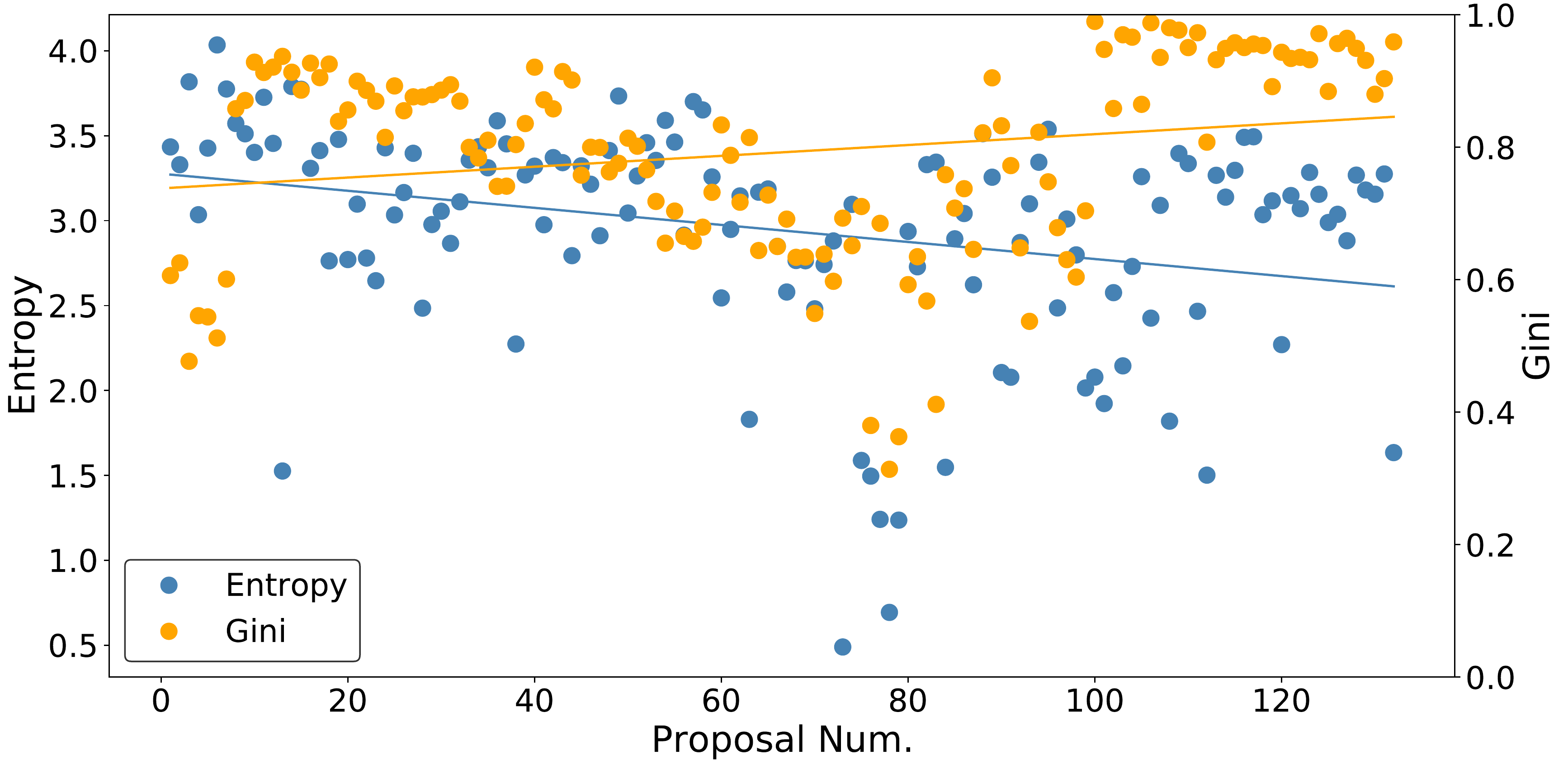}
  \caption{CompoundDAO Decentralization Level Overtime}\label{fig:decen-level1}
\end{subfigure}\hfill
\begin{subfigure}{0.45\textwidth}
  \includegraphics[width=\linewidth]{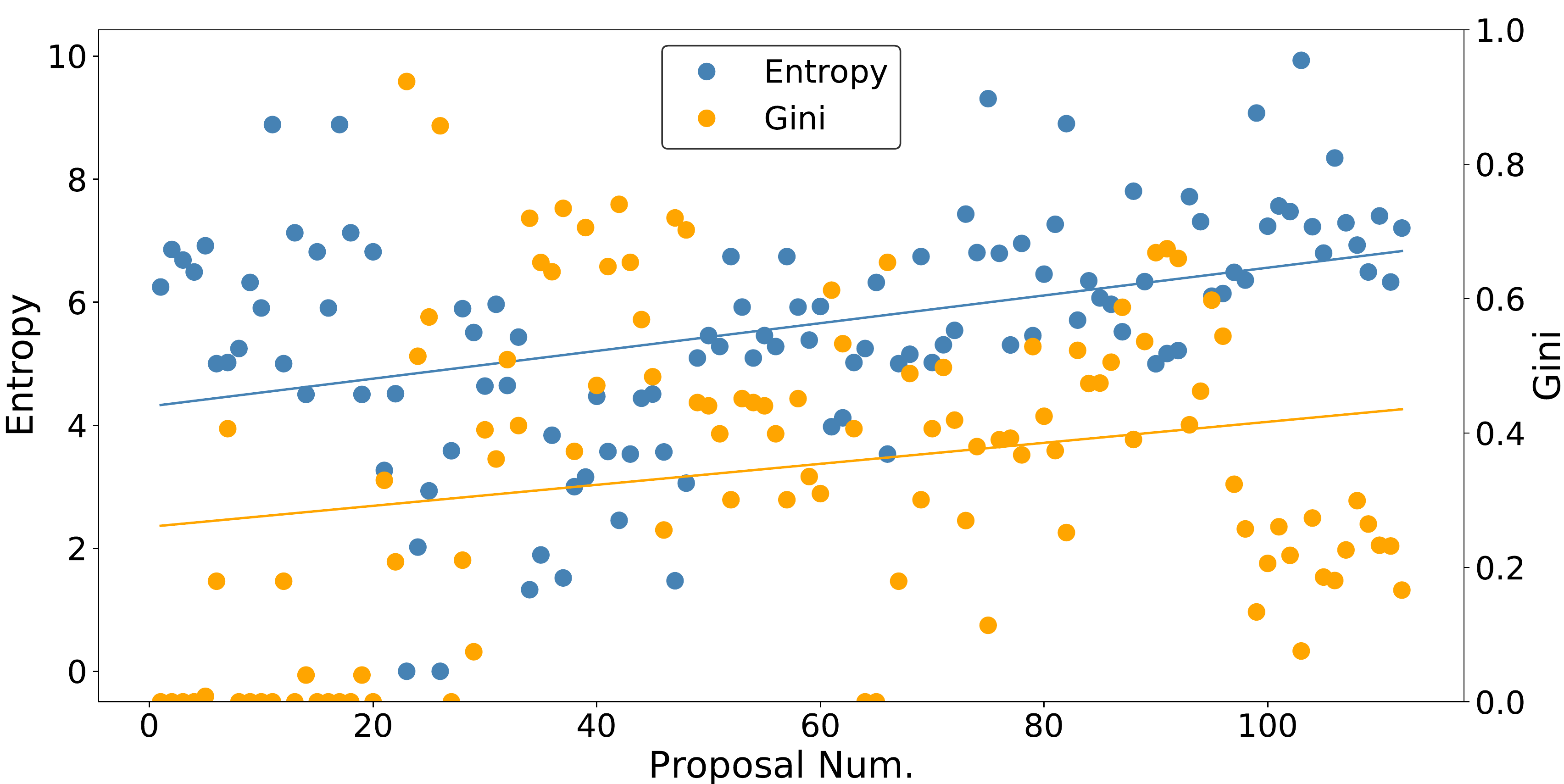}
  \caption{Proof of Humanity: Decentralization level over time }
  \label{fig:decen-level2}
\end{subfigure}
\caption{Voting Power Evaluation of decentralization level by Entropy, Gini coefficient, and Nakamoto showed significantly poor decentralization for Compound, BitDAO, Krasehouse, MolochDAO, and dxDAO while the highest decentralization level for Proof of Humanity. Unlike many DAO systems including CompoundDAO, the decentralization level of PoH is in increasing trend over time.}
\label{cluster2}
\end{figure*}
  
Next, we evaluated decentralization level of 10 DAOs in terms of the voting power. To do this, we used three metrics: Entropy, the Gini and Nakamoto coefficient. Entropy can be generally used to assess randomness and degrees of freedom for a given system. Mathematically, the metric for given a series of numbers $x (={x_i})$ is defined as follows: 
$$\vspace{-2mm}Entropy(x)=\sum_{i}log_2\left(\frac{x_i}{\sum_{i}x_i}\right)*\frac{x_i}{\sum_{i}x_i}.\vspace{-2mm}$$ 
If the value of entropy is high, it implies a DAO has a high degree of randomness and freedom, which is related to a high level of decentralization. The Gini coefficient used to measure the wealth inequality traditionally can be applied to assess the voting power inequality among voters in our analysis. The metric for given a series of numbers x is defined as follows: 
$$\vspace{-2mm}Gini(x)=\frac{\sum_{i,j}|x_i-x_j|}{2 |x| \sum_{i}x_i},$$ 
where $|x|$ indicates the length of the number series. The minimum value of the Gini coefficient as 0 would represent the full equality among voters, while the maximum value of 1 would represent the full power bias towards one voter. We calculated Entropy and Gini based on the token weight of voters and their voting scores, which, in this paper, we define as \textit{``token weight$\times$the number of proposals that a voter participated in''} to reflect not only the token weight but also a participation rate. Lastly, the Nakamoto coefficient indicates the minimum number of entities that can subvert the system. In the voting context, the entities can accept or reject the proposal in accordance with their interests, even if it may be against most users. Therefore, if power is significantly biased towards only a few voters, the DAO cannot be considered to attain true democracy, which is one of the main goals of DAOs. We calculated the Nakamoto coefficient based on the token weight of voters.    

From Table~\ref{tab:decen-level}, we can see almost all DAOs suffer from poor decentralization. In particular, Compound, BitDAO, Krasehouse, LivePeer, Meta Gamma Delta, MolochDAO, and dxDAO showed significantly poor decentralization, where their Gini coefficients were above 0.9 due to a extremely large power inequliaty or (and) the Entropy values were low due to a low number of participating voters. On the other hand,  Proof of Humanity showed the highest decentralization level among our target DAOs. In fact, it uses one-person-one-vote protocol unlike the popular protocol that the voting power is proportional to token amounts holded by a voter. However, note that our result analysis does not necessarily mean that one-person-one-vote protocol is better than the other protocol using the token capital concept; even though Meta Gamma Delta also uses a similar protocol to Proof of Humanity, it currently suffers from the low number of active voters. 

\begin{tcolorbox}[width=\linewidth, colback=white!95!black, boxrule=0.5pt, left=2pt,right=2pt,top=1pt,bottom=1pt]
\stepcounter{observation}
{\bf Takeaways \arabic{observation}:}
{In many DAOs, richer token holders tend to participate more actively in the voting process, which aggravates the decentralization of the voting process. Moreover, almost all DAOS that we analyzed are suffering from poor decentralization. 
}

\end{tcolorbox}

\subsubsection{Decentralization Dynamics Overtime} 

We also present the decentralization dynamics over time. Table~\ref{tab:decen-over} in Appendix summarizes the linear regression results of 10 DAOs, where we mark significantly negative and positive decentralization trends with red and blue, respectively. Overall, we can see that many DAOs (CompoundDAO, BitDAO, AssangeDAO, BanklessDAO, and KrauseHouse) show a significantly negative decentralization trend: the decreasing decentralization level over time (Figure~\ref{fig:decen-level1}). On the other hand, Proof of Humanity and MGD have a significantly positive decentralization trend: the increasing decentralization level over time (Figure~\ref{fig:decen-level2}). Note that, according to Table~\ref{tab:decen-level}, in the Proof of Humanity system, a biased participation rate of voters is more a matter rather than a biased token weight distribution  in terms of a decentralization level; entropy and the Gini coefficient for voters’ token weight are high and low, respectively. Therefore, we focus on the linear regression result for Entropy rather than that for the Gini coefficient shown in Table~\ref{tab:decen-over}. One can find that both DAOs, Proof of Humanity and MGD, are a system adopting the one-person-one-vote protocol. As a result, the level of decentralization has got better over time in the one-person-one-vote protocol. 


\begin{tcolorbox}[width=\linewidth, colback=white!95!black, boxrule=0.5pt, left=2pt,right=2pt,top=1pt,bottom=1pt]
\stepcounter{observation}
{\bf Takeaways \arabic{observation}:}
{The decentralization dynamics analysis shows that decentralization has been gradually aggravated in a voting protocol based on token capital, while it has improved over time in the one-person-one-vote protocol.
}

\end{tcolorbox}

\vspace{-2mm}
\section{RQ3: Autonomy of DAOs}
\vspace{-2mm}
\label{section6}

For the DAO autonomy analysis, we explore (1) whether a DAO system can execute arbitrary transactions, (2) whether it relies on third parties after transaction execution, and (3) whether proposals have been canceled after being successful. 
\vspace{-2mm}
\subsection{Capability of Arbitrary Transaction Execution}
\vspace{-2mm}
Arbitrary transaction execution refers to the capability of a DAO to fulfill various proposals, regardless of their specific requirements. Without this capability, a DAO's functionality is limited, requiring additional intervention to execute certain transactions on-chain, like transferring funds to an EOA\footnote{Externally Owned Account (EOA) refers to addresses that are managed by a private key and not a smart contract code.} to carry out proposal execution. This would decrease the DAO autonomy. To achieve arbitrary transaction execution, DAOs should interact with any contract 
by passing call data through the call function in Solidity~\cite{das2019fastkitten}. 

To determine the capability of a DAO to execute arbitrary transactions and related methods, 
we first thoroughly examined the smart contracts listed on the DAO's website or Github repository. 
Specifically, we searched for code snippets that enable arbitrary transaction execution. 

Our study indicates that 10 DAOs have protocols for executing arbitrary transactions, albeit with varying autonomy levels. Table~\ref{Tab:Tab-auto-met-1} in Appendix summarizes the protocols used in executing arbitrary transactions.
Most DAOs that employ on-chain voting, including CompoundDAO, Meta Gamma Delta, Moloch, and dxDAO, exhibit an entire pipeline for voting, consensus building, and execution. Consequently, we find such DAOs to be most autonomous, as the transaction sequence is determined upon proposal submission and immutable thereafter. Conversely, DAOs employing off-chain voting necessitate a means of transposing the off-chain consensus onto the on-chain ledger, such as multisig wallet to issue on-chain transactions. Most off-chain voting DAOs, including AssangeDAO, BitDAO, Banklass, and KrauseHouse, and LivePeer use a multisig wallet to issue on-chain transactions. Multisig lowers decentralization and autonomy levels, as a small group wields controls over the on-chain actions to be executed, however, eases the implementation and maintenance. 
Alternately, on-chain transactions to fulfill approved proposals can be crafted and refined by community members through a system of checks and balances involving monetary stakes.
Kleros Governor contract~\cite{kleros} exemplifies this approach, which is adopted by Proof of Humanity. Although this method is more decentralized than a multisig wallet, it still limits autonomy, as multiple parties must verify 
the proposed transactions list, which, but, would delay execution. 
Figures~\ref{fig:code-snippet-1} and \ref{fig:code-snippet-2} in Appendix provide examples of code snippets for different types of arbitrary transactions, including on-chain
and multisig. 

\begin{tcolorbox}[width=\linewidth, colback=white!95!black, boxrule=0.5pt, left=2pt,right=2pt,top=1pt,bottom=1pt]
\stepcounter{observation}
{\bf Takeaways \arabic{observation}:}
{
DAOs vary in their approach to executing arbitrary transactions, with fully on-chain pipeline being the most autonomous.
Multisig approach trades autonomy for simplicity, while Kleros contract also lacks autonomy even though it increases decentralization. 
}
\end{tcolorbox}

\vspace{-3mm}

\vspace{-2mm}
\subsection{Third Party Dependency in Proposals}
\vspace{-2mm}
In DAOs, the reliance on external entities to fulfill proposals after the voting period is commonly known as third-party dependencies.
This reliance can imply lower DAO autonomy (please see Sections~\ref{subsec:rw-autonomous} and \ref{interview}), introducing uncontrollable external factors.
We investigate how current DAOs utilizing third-party services handle and minimize associated risks. Analyzing the relationship between DAOs and third parties can not only aid in evaluating autonomy but also provide insight into how DAOs currently benefit from third-party services, minimizing risk exposure. 
Third party dependent proposals are defined as those relying on other entities apart from the DAO itself to fulfill a part of the proposal after passing the voting period.
These proposals include payment for future services or delegating decision-making power to a small group. 
To investigate the utilization of third-party services, we analyzed both the content of proposals and associated execution data. Our evaluation process involved assessing the primary objectives of proposals to identify any post-execution intervention or deviation from the proposal's promises that could potentially negate the effectiveness or have a detrimental effect on DAOs.


Table~\ref{tab:3rd-party} presents the percentage of third party dependent proposals for nine different DAOs. dxDAO is left out of the analysis due to the large number of proposals making it infeasible to conduct the manual analysis. Of the 9 DAOs examined, only MetaGammaDelta DAO proposals do not rely on third-party services after execution because they are typically requests for membership or funding, with no return promises. The DAO with the highest percentage of third party dependent proposals is KrauseHouse ($74.81\%$). Many KrauseHouse proposals involve funding requests from third parties with commitments of future returns in the form of working hours.
While third-party reliance following proposal execution can be beneficial in allowing a DAO to achieve various goals, particularly those that are off-chain, DAOs should take measures to minimize the risks associated with third-party services. Our analysis identified strategies used by DAOs to mitigate these risks--(a) using stream payments instead of upfront lump sum payments, (b) using milestone-based payments where funds are paid based on completed tasks, and (c) assigning a committee with a multisig wallet to manage funds allocated for third-party.

\begin{tcolorbox}[width=\linewidth, colback=white!95!black, boxrule=0.5pt, left=2pt,right=2pt,top=1pt,bottom=1pt]
\stepcounter{observation}
{\bf Takeaways \arabic{observation}:}
{
Most DAOs require services from third parties to operate with common case is employment for some future services. 
Despite such necessity of third parties in DAOs, there are many strategies to minimize the risk of third-party reliance.
}
\end{tcolorbox}
\vspace{-2mm}
\subsection{Proposals Canceled after Voting Ends}
\vspace{-2mm}
Subsequently, we examined proposals that were accepted but not implemented, referred to as \emph{``canceled proposals.''} 
This measure serves as a proxy for evaluating whether community consensus is respected and whether a DAO is truly autonomous, i.e., whether a small group of individuals can prevent a DAO action from being carried out. 
We analyzed events from the smart contracts of on-chain DAOs, except for LivePeer. For off-chain DAOs 
we manually examined data from Snapshot and Discourse to determine if the proposals have been successfully implemented.

Meta Gamma Delta, MolochDAO, and dxDAO have a ``binding'' consensus, indicating proposals are executed after acceptance through a smart contract, with no cancellation mechanism after the proposal has passed. Only 2 on-chain DAOs, CompoundDAO and Livepeer have a formal cancellation mechanism after the proposal has passed (Table~\ref{tab:canceled} in Appendix). In CompoundDAO, proposal cancellation can be done by the proposal author or by anyone if author's governance token balance drops below proposal threshold.
The proposals that were canceled after the voting period ends in CompoundDAO are mostly those overwritten by another proposal because of parameter/transaction issues~\cite{compound} and ones for a temporary bug fix ~\cite{compound1}. In the case of LivePeer, DAO requires a multisig to issue the transaction, and not everything is done on-chain. Thus, after the consensus is reached, multisig still can \emph{``cancel''} the proposal if they wish. For the remaining off-chain DAOs, proposals can technically be censored as off-chain consensus is not guaranteed to be brought on-chain. However, some off-chain DAOs still hold a binding consensus ideology. 
While the percentage of proposals canceled is an important metric for autonomy, difficulty arises when empirically calculating this metric due to the limited transparency of the proposal status after consensus is reached in many off-chain DAOs.

\begin{tcolorbox}[width=\linewidth, colback=white!95!black, boxrule=0.5pt, left=2pt,right=2pt,top=1pt,bottom=1pt]
\stepcounter{observation}
{\bf Takeaways \arabic{observation}:}
{
Our data on the number of proposals canceled after voting ends suggests that DAOs tend to follow proposal binding ideology even if it is not guaranteed programmatically. A high number of unknown status proposals suggests that there is a need for proposal transparency tools.
}
\end{tcolorbox}

\begin{table}[]

\begin{tabular}{lll}
\scriptsize

Name              & 1 (not reliance) & 0 (reliance) \\
\hline
AssangeDAO        & 81.81\%          & 18.18\%      \\
CompoundDAO       & 83.94\%          & 16.06\%      \\
BitDAO            & 72.22\%          & 28.78\%      \\
Proof of Humanity & 86.54\%          & 13.46\%      \\
Bankless DAO      & 70.59\%          & 29.41\%      \\
KrauseHouse       & 25.19\%          & 74.81\%      \\
LivePeer          & 85.71\%          & 14.29\%      \\
Meta Gamma Delta  & 100\%            & 0\%          \\
MolochDAO         & 84.84\%          & 15.15\%    \\
\hline
\end{tabular}
\caption{Percentage of proposals that relies on third-party services after execution. Rows are each DAOs where 0 means relying on a third party and 1 is not reliance}
\label{tab:3rd-party}
\end{table}
\vspace{-2mm}
\section{Discussion}
\vspace{-2mm}
\label{section7}
In this section, we discuss the prevailing governance structure of DAOs and the degree to which they realize their envisioned goals. Based on the current outcomes of DAOs, we also provide \emph{governance and design implications} of DAOs, and identify areas warranting further investigation.

\vspace{-2mm}
\subsection{Recap of Main Findings} 
\vspace{-2mm}
DAO represents a significant artifact for comprehending emerging forms of algorithmic authority, exploring practical modes of governance for autonomous and decentralized systems, and understanding the ways that designing incentives and modeling can fail~\cite{dupont2017experiments}. In our interview, DAO experts indicated three metrics including, token distribution, voter participation, and geographical distribution for decentralization, while some of these were evaluated in our subsequent empirical analysis to understand the impact in the decentralization level. Our constructed voting network graph for each DAO and calculated clustering and connectivity metrics show that there is a higher degree of voting pattern polarization of large token holders relative to other DAO members. We also found a significant positive correlation between the proposal’s author token balance and the proposal's success in some DAOs, which may be a bad sign in DAOs where voting power is traded on the secondary market. To quantify decentralization, we applied entropy, the Gini coefficient, and the Nakamoto coefficient of the token distribution and the voting participation rate of voters. We found that DAOs that allow governance tokens to be bought from the secondary market are less decentralized over time, contradicting the expectation of a DAO becoming more decentralized over time, while those adopting the one-person-one-vote protocol have increased decentralization.

In terms of autonomy, we explored the dependency of a DAO on external entities and the degree of intervention a DAO needs to carry out its actions. We found that many offchain DAOs use a Multisig wallet to translate the consensus reached offchain to onchain actions, resulting in lower autonomy.  Kleros Governor is an interesting alternative, but the method can lack autonomy as it relies on multi-step checks done after consensus.
The portion of proposals relying on third-party services after execution varies from 0-75\% among the DAOs studied. Another metric calculated in this study is the number of proposals canceled after voting ends. This metric is particularly important as it shows to what degree the community consensus is respected, however this analysis is constrained by the limited transparency of offchain DAOs. The large number of such proposals call for the need for more transparent proposal tracking tools.

In summary, we provide an overall characterization of governance structures present in DAOs and assess their level of success. Subsequently, we delve into the extent to which the observed outcomes can be attributed to the design choices made in the DAOs, such as the role of marketing factors versus the impact of the governance process design. 
\vspace{-2mm}
\subsection{Design Implication}
\vspace{-2mm}
\noindent\textbf{Incentive mechanism: separation between voting power and monetary value.} 
One of the most critical points that enable healthy and robust DAOs is facilitating equitable participation of numerous entities which necessitates proper incentive systems~\cite{argon}. Our paper confirmed that 
biased voting power distribution or low participation rate in many existing DAOs can be attributed to a lack of proper incentive system
~\cite{kaal2021decentralized}. Whereas DAO experts also suggested a separation between voting power and monetary value
might be needed to achieve better decentralization. 
Although the optimal incentive designs for decentralization are unclear, DAO practitioners can consider a handful of design choices to help address this issue. First, DAO practitioners can choose to have nontransferable/non-monetary vote tokens. They can consider social rewards like reputation, such as, reputation-based voting, and contribution-based voting, one person one vote. Some studies~\cite{liu2022user,guidi2020graph} showed that social rewards could be effective in promoting the engagement of users in the DAO context. Alternatively, DAOs with existing transferable voting tokens can choose to adopt a quadratic voting scheme, making the purchase of a unit voting power increasingly expensive as users hold more tokens.

\noindent\textbf{AI-based intervention protocol with bubble-bursting fact-checks.} 
Fact-checking and moderation have been devised to encourage users to explore a wider range of viewpoints, transcend social media echo chambers, and ultimately decrease political and cultural polarization~\cite{wardle2017information,nekmat2020nudge}. 
In the context of DAOs, we have observed an unequal polarization among proposal voters, where whales exhibit homogeneous voting behavior among themselves, failing to represent the wider population. Experts have attributed this phenomenon to the lack of technical understanding of proposals of voters, leading them to follow the voting patterns of their trusted key opinion leaders. 
To address this polarization and echo chamber effect, an intervention protocol (in the form of a ML based recommender model) could be developed to enhance the credibility of information and prioritize fact-checking messages. Such a model would nudge DAO holders to learn more about proposals, thereby allow informed decision-making during voting rather than relying on the KOLs. This intervention protocol can allow diverse proposal content consumption during the soft proposal stage for both whales and grassroot voters.

\noindent\textbf{Onchain proposal pipeline for autonomy \& ux.} 
Despite the benefits of offchain tools like Snapshot's gasless and verifiable voting platform, the manual translation of offchain consensus to onchain actions poses a bottleneck for decentralization and autonomy as the transactions are often manually crafted by a small group of people in most DAOs.
One solution to this could be moving the entire proposal pipeline onchain. While on-chain platforms encourage higher transparency and autonomy compared to Snapshot and Multisig, current on-chain systems of contracts are more difficult to deploy and maintain. Tools that allow DAO practitioners and users to have better user experience in deploying, maintaining, and interacting with fully on-chain DAO system of contracts can help achieve the best of both worlds. With features like stream payment and binding consensus being readily available onchain, this reduces the reliance on third-party services. In addition, in DAOs that suffer from long delays, being fully onchain opens up the possibility for any user to help issue the transaction to execute a proposal. 
Monetary incentives could also further incentivize users to execute DAO actions promptly. Of course, how to incentivize users in DAO would remain an open problem. 
Layer 2 solutions can help lower the gas cost for users, costing only a few cents to participate.


\noindent\textbf{Fully automated tracking tools for proposal transparency.} In examining third party dependent proposals, many proposals lack transparency during the execution period. This results in a handful of proposals having unknown status after passing the voting period. Reaching consensus alone is not enough to ensure autonomy. The resulting DAO actions must be easily verifiable by DAO members. At the moment, there are several tools employed by DAOs to publicly post and interpret DAOs action, such as Notion. However, there is a lack of adoption for fully automated tool to track proposal status from the start until after consensus is reached. In DAOs that rely heavily on employing third parties, fully automated tools that help track the progress of the task, end result, and payment are essential for DAO members to make an informed decision on compensation and future employments.

\bibliographystyle{IEEEtran}
\bibliography{sample}

\begin{thebibliography}{10}
\providecommand{\url}[1]{#1}
\csname url@samestyle\endcsname
\providecommand{\newblock}{\relax}
\providecommand{\bibinfo}[2]{#2}
\providecommand{\BIBentrySTDinterwordspacing}{\spaceskip=0pt\relax}
\providecommand{\BIBentryALTinterwordstretchfactor}{4}
\providecommand{\BIBentryALTinterwordspacing}{\spaceskip=\fontdimen2\font plus
\BIBentryALTinterwordstretchfactor\fontdimen3\font minus
  \fontdimen4\font\relax}
\providecommand{\BIBforeignlanguage}[2]{{%
\expandafter\ifx\csname l@#1\endcsname\relax
\typeout{** WARNING: IEEEtran.bst: No hyphenation pattern has been}%
\typeout{** loaded for the language `#1'. Using the pattern for}%
\typeout{** the default language instead.}%
\else
\language=\csname l@#1\endcsname
\fi
#2}}
\providecommand{\BIBdecl}{\relax}
\BIBdecl

\bibitem{messari}
Accessed on 2023, \url{https://messari.io/}.

\bibitem{dwivedi2021legally}
V.~Dwivedi, V.~Pattanaik, V.~Deval, A.~Dixit, A.~Norta, and D.~Draheim,
  ``Legally enforceable smart-contract languages: A systematic literature
  review,'' \emph{ACM Computing Surveys (CSUR)}, vol.~54, no.~5, pp. 1--34,
  2021.

\bibitem{esteves2023potential}
N.~Est{\`e}ves, A.~Wernick, and S.~Carls, ``The potential of follow-on
  innovation financing instruments to support a sustainable transition,''
  \emph{Intellectual Property Rights in the Post Pandemic World: an Integrated
  Framework of Sustainability, Innovation and Global Justice. Edward Elgar},
  pp. 22--07, 2023.

\bibitem{chaisse2022tokenised}
J.~Chaisse and J.~Kirkwood, ``Tokenised funding and initial litigation
  offerings: the new kids putting third-party funding on the block,'' \emph{Law
  and Financial Markets Review}, pp. 1--23, 2022.

\bibitem{coindesk}
Accessed on 2023,
  \url{https://www.coindesk.com/learn/what-is-a-governance-token/ }.

\bibitem{hassan2021decentralized}
S.~Hassan and P.~De~Filippi, ``Decentralized autonomous organization,''
  \emph{Internet Policy Review}, vol.~10, no.~2, pp. 1--10, 2021.

\bibitem{hackernoon}
Accessed on 2023,
  \url{https://hackernoon.com/what-are-the-benefits-of-dao-governance-16e8b66f17e7
  }.

\bibitem{zachariadis2019governance}
M.~Zachariadis, G.~Hileman, and S.~V. Scott, ``Governance and control in
  distributed ledgers: Understanding the challenges facing blockchain
  technology in financial services,'' \emph{Information and Organization},
  vol.~29, no.~2, pp. 105--117, 2019.

\bibitem{zwitter2020decentralized}
A.~Zwitter and J.~Hazenberg, ``Decentralized network governance: blockchain
  technology and the future of regulation,'' \emph{Frontiers in Blockchain},
  vol.~3, p.~12, 2020.

\bibitem{zwitter2020governance}
------, ``Governance, blockchain, cyberspace: How technology implies normative
  power and regulation,'' \emph{Blockchain, Law and Governance, Springer,
  Forthcoming}, 2020.

\bibitem{chohan2017decentralized}
U.~W. Chohan, ``The decentralized autonomous organization and governance
  issues,'' \emph{Available at SSRN 3082055}, 2017.

\bibitem{mehar2019understanding}
M.~I. Mehar, C.~L. Shier, A.~Giambattista, E.~Gong, G.~Fletcher, R.~Sanayhie,
  H.~M. Kim, and M.~Laskowski, ``Understanding a revolutionary and flawed grand
  experiment in blockchain: the dao attack,'' \emph{Journal of Cases on
  Information Technology (JCIT)}, vol.~21, no.~1, pp. 19--32, 2019.

\bibitem{liu2021technology}
L.~Liu, S.~Zhou, H.~Huang, and Z.~Zheng, ``{From technology to society: An
  overview of blockchain-based DAO},'' \emph{IEEE Open Journal of the Computer
  Society}, vol.~2, pp. 204--215, 2021.

\bibitem{dhillon2017dao}
V.~Dhillon, D.~Metcalf, M.~Hooper, V.~Dhillon, D.~Metcalf, and M.~Hooper, ``The
  dao hacked,'' \emph{blockchain enabled applications: Understand the
  blockchain Ecosystem and How to Make it work for you}, pp. 67--78, 2017.

\bibitem{morrison2020dao}
R.~Morrison, N.~C. Mazey, and S.~C. Wingreen, ``The dao controversy: the case
  for a new species of corporate governance?'' \emph{Frontiers in Blockchain},
  vol.~3, p.~25, 2020.

\bibitem{dupont2017experiments}
Q.~DuPont, ``{Experiments in algorithmic governance: A history and ethnography
  of ``The DAO,'' a failed decentralized autonomous organization},'' in
  \emph{Bitcoin and beyond}.\hskip 1em plus 0.5em minus 0.4em\relax Routledge,
  2017, pp. 157--177.

\bibitem{mosley2022towards}
L.~Mosley, H.~Pham, X.~Guo, Y.~Bansal, E.~Hare, and N.~Antony, ``Towards a
  systematic understanding of blockchain governance in proposal voting: A dash
  case study,'' \emph{Blockchain: Research and Applications}, p. 100085, 2022.

\bibitem{trisetyarso2019crypto}
A.~Trisetyarso, W.~Suparta, C.-H. Kang, B.~S. Abbas \emph{et~al.},
  ``Crypto-governance in stock exchanges: Towards efficient and self-regulated
  trading system,'' in \emph{2019 International Conference on contemporary
  Computing and Informatics (IC3I)}.\hskip 1em plus 0.5em minus 0.4em\relax
  IEEE, 2019, pp. 192--197.

\bibitem{myeong2019administrative}
S.~Myeong and Y.~Jung, ``Administrative reforms in the fourth industrial
  revolution: the case of blockchain use,'' \emph{Sustainability}, vol.~11,
  no.~14, p. 3971, 2019.

\bibitem{barbosa2018cryptocurrencies}
A.~C. Barbosa, T.~A. Oliveira, and V.~N. Coelho, ``Cryptocurrencies for smart
  territories: an exploratory study,'' in \emph{2018 International Joint
  Conference on Neural Networks (IJCNN)}.\hskip 1em plus 0.5em minus
  0.4em\relax IEEE, 2018, pp. 1--8.

\bibitem{chatterjee2019probabilistic}
K.~Chatterjee, A.~K. Goharshady, and A.~Pourdamghani, ``Probabilistic smart
  contracts: Secure randomness on the blockchain,'' in \emph{2019 IEEE
  International Conference on Blockchain and Cryptocurrency (ICBC)}.\hskip 1em
  plus 0.5em minus 0.4em\relax IEEE, 2019, pp. 403--412.

\bibitem{ciatto2018blockchain}
G.~Ciatto, R.~Calegari, S.~Mariani, E.~Denti, and A.~Omicini, ``From the
  blockchain to logic programming and back: Research perspectives.'' in
  \emph{WOA}, 2018, pp. 69--74.

\bibitem{tse2020decentralised}
N.~Tse, ``Decentralised autonomous organisations and the corporate form,''
  \emph{Victoria U. Wellington L. Rev.}, vol.~51, p. 313, 2020.

\bibitem{ellul2019blockchain}
J.~Ellul and G.~Pace, ``Blockchain and the common good reimagined,''
  \emph{arXiv preprint arXiv:1910.14415}, 2019.

\bibitem{bowles2012democracy}
S.~Bowles and H.~Gintis, \emph{Democracy and capitalism: Property, community,
  and the contradictions of modern social thought}.\hskip 1em plus 0.5em minus
  0.4em\relax Routledge, 2012.

\bibitem{buterin2014next}
V.~Buterin \emph{et~al.}, ``A next-generation smart contract and decentralized
  application platform,'' \emph{white paper}, vol.~3, no.~37, pp. 2--1, 2014.

\bibitem{williamson2002theory}
O.~E. Williamson, ``The theory of the firm as governance structure: from choice
  to contract,'' \emph{Journal of economic perspectives}, vol.~16, no.~3, pp.
  171--195, 2002.

\bibitem{shaw2002public}
J.~S. Shaw, ``Public choice theory,'' \emph{The concise encyclopedia of
  economics}, 2002.

\bibitem{nurmi1999voting}
H.~Nurmi, \emph{Voting paradoxes and how to deal with them}.\hskip 1em plus
  0.5em minus 0.4em\relax Springer Science \& Business Media, 1999.

\bibitem{benoit2000gibbard}
J.-P. Beno{\i}t, ``The gibbard--satterthwaite theorem: a simple proof,''
  \emph{Economics Letters}, vol.~69, no.~3, pp. 319--322, 2000.

\bibitem{satterthwaite1975strategy}
M.~A. Satterthwaite, ``Strategy-proofness and arrow's conditions: Existence and
  correspondence theorems for voting procedures and social welfare functions,''
  \emph{Journal of economic theory}, vol.~10, no.~2, pp. 187--217, 1975.

\bibitem{faraj2006coordination}
S.~Faraj and Y.~Xiao, ``Coordination in fast-response organizations,''
  \emph{Management science}, vol.~52, no.~8, pp. 1155--1169, 2006.

\bibitem{fish1988quilt}
R.~S. Fish, R.~E. Kraut, and M.~D. Leland, ``Quilt: A collaborative tool for
  cooperative writing,'' in \emph{Proceedings of the ACM SIGOIS and IEEECS
  TC-OA 1988 conference on Office information systems}, 1988, pp. 30--37.

\bibitem{stokols2008ecology}
D.~Stokols, S.~Misra, R.~P. Moser, K.~L. Hall, and B.~K. Taylor, ``The ecology
  of team science: understanding contextual influences on transdisciplinary
  collaboration,'' \emph{American journal of preventive medicine}, vol.~35,
  no.~2, pp. S96--S115, 2008.

\bibitem{baninemeh2023decision}
E.~Baninemeh, S.~Farshidi, and S.~Jansen, ``A decision model for decentralized
  autonomous organization platform selection: Three industry case studies,''
  \emph{Blockchain: Research and Applications}, p. 100127, 2023.

\bibitem{wright2021measuring}
S.~A. Wright, ``Measuring dao autonomy: Lessons from other autonomous
  systems,'' \emph{IEEE Transactions on Technology and Society}, vol.~2, no.~1,
  pp. 43--53, 2021.

\bibitem{santana2022blockchain}
C.~Santana and L.~Albareda, ``Blockchain and the emergence of decentralized
  autonomous organizations (daos): An integrative model and research agenda,''
  \emph{Technological Forecasting and Social Change}, vol. 182, p. 121806,
  2022.

\bibitem{sae2018taxonomy}
S.~International, ``Taxonomy and definitions for terms related to driving
  automation systems for on-road motor vehicles,'' \emph{SAE international},
  vol. 4970, no. 724, pp. 1--5, 2018.

\bibitem{elhannouny2019off}
E.~M. Elhannouny and D.~Longman, ``Off-road vehicles research workshop: Summary
  report,'' Argonne National Lab.(ANL), Argonne, IL (United States), Tech.
  Rep., 2019.

\bibitem{clough2002metrics}
B.~T. Clough, ``Metrics, schmetrics! how the heck do you determine a uav's
  autonomy anyway,'' Air Force Research Lab Wright-Patterson AFB OH, Tech.
  Rep., 2002.

\bibitem{federal2015operation}
F.~A. ADMINISTRATION.(FAA), \emph{Operation and Certification of Small Unmanned
  Aircraft Systems}.\hskip 1em plus 0.5em minus 0.4em\relax LULU COM, 2015.

\bibitem{huang2005autonomy}
H.-M. Huang, K.~Pavek, J.~Albus, and E.~Messina, ``Autonomy levels for unmanned
  systems (alfus) framework: An update,'' in \emph{Unmanned Ground Vehicle
  Technology VII}, vol. 5804.\hskip 1em plus 0.5em minus 0.4em\relax SPIE,
  2005, pp. 439--448.

\bibitem{heckman1998liability}
C.~Heckman and J.~O. Wobbrock, ``Liability for autonomous agent design,'' in
  \emph{Proceedings of the second international conference on Autonomous
  agents}, 1998, pp. 392--399.

\bibitem{dowling2000intelligent}
C.~Dowling, ``Intelligent agents: some ethical issues and dilemmas,''
  \emph{Proc. AIC 2000}, pp. 28--32, 2000.

\bibitem{ethereum}
Accessed on 2023,
  \url{https://blog.ethereum.org/2014/05/06/daos-dacs-das-and-more-an-incomplete-terminology-guide}.

\bibitem{pagallo2017automation}
U.~Pagallo \emph{et~al.}, ``From automation to autonomous systems: A legal
  phenomenology with problems of accountability,'' in \emph{IJCAI International
  Joint Conference on artificial intelligence}.\hskip 1em plus 0.5em minus
  0.4em\relax International Joint Conferences on Artificial Intelligence, 2017,
  pp. 17--23.

\bibitem{dyrkolbotn2017classifying}
S.~K. Dyrkolbotn, T.~Pedersen, and M.~Slavkovik, ``Classifying the autonomy and
  morality of artificial agents.'' \emph{CARe-MAS@ PRIMA}, vol. 2017, pp.
  67--83, 2017.

\bibitem{haupt2019artificial}
C.~E. Haupt, ``Artificial professional advice,'' \emph{Yale JL \& Tech.},
  vol.~21, p.~55, 2019.

\bibitem{ladner2019local}
A.~Ladner, N.~Keuffer, H.~Baldersheim, N.~Hlepas, P.~Swianiewicz, K.~Steyvers,
  C.~Navarro, A.~Ladner, N.~Keuffer, H.~Baldersheim \emph{et~al.}, ``The local
  autonomy index (lai),'' \emph{Patterns of local autonomy in Europe}, pp.
  213--254, 2019.

\bibitem{ryan2019brick}
R.~M. Ryan and E.~L. Deci, ``Brick by brick: The origins, development, and
  future of self-determination theory,'' in \emph{Advances in motivation
  science}.\hskip 1em plus 0.5em minus 0.4em\relax Elsevier, 2019, vol.~6, pp.
  111--156.

\bibitem{kramerresponsible}
A.~S. Kramer, ``Responsible financial innovation act: Proposed tax and
  reporting for digital assets.''

\bibitem{citydao}
Accessed on 2023, \url{https://www.citydao.io/topic/weekly-news }.

\bibitem{Etherscan}
Accessed on 2023, \url{https://etherscan.io}.

\bibitem{gnosis}
Accessed on 2022, \url{https://docs.snapshot.org/gnosis-safe}.

\bibitem{daohaus}
Accessed on 2022, \url{https://daohaus.substack.com/}.

\bibitem{petrochilos2013using}
D.~Petrochilos, A.~Shojaie, J.~Gennari, and N.~Abernethy, ``Using random walks
  to identify cancer-associated modules in expression data,'' \emph{BioData
  mining}, vol.~6, pp. 1--25, 2013.

\bibitem{meghanathan2016assortativity}
N.~Meghanathan, ``Assortativity analysis of real-world network graphs based on
  centrality metrics.'' \emph{Comput. Inf. Sci.}, vol.~9, no.~3, pp. 7--25,
  2016.

\bibitem{comp}
Accessed on 2023, \url{https://www.comp.xyz/}.

\bibitem{das2019fastkitten}
P.~Das, L.~Eckey, T.~Frassetto, D.~Gens, K.~Host{\'a}kov{\'a}, P.~Jauernig,
  S.~Faust, and A.-R. Sadeghi, ``$\{$FastKitten$\}$: Practical smart contracts
  on bitcoin,'' in \emph{28th USENIX Security Symposium (USENIX Security 19)},
  2019, pp. 801--818.

\bibitem{kleros}
Accessed on 2023, \url{
  https://github.com/kleros/kleros/blob/master/contracts/kleros/KlerosGovernor.sol}.

\bibitem{compound}
Accessed on 2023, \url{https://compound.finance/governance/proposals/99}.

\bibitem{compound1}
Accessed on 2023, \url{https://compound.finance/governance/proposals/63}.

\bibitem{argon}
Accessed on 2023,
  \url{https://blog.aragon.org/incentive-design-tooling-for-daos/}.

\bibitem{kaal2021decentralized}
W.~A. Kaal, ``A decentralized autonomous organization (dao) of daos,''
  \emph{Available at SSRN 3799320}, 2021.

\bibitem{liu2022user}
Z.~Liu, Y.~Li, Q.~Min, and M.~Chang, ``User incentive mechanism in
  blockchain-based online community: An empirical study of steemit,''
  \emph{Information \& Management}, p. 103596, 2022.

\bibitem{guidi2020graph}
B.~Guidi, A.~Michienzi, and L.~Ricci, ``A graph-based socioeconomic analysis of
  steemit,'' \emph{IEEE Transactions on Computational Social Systems}, vol.~8,
  no.~2, pp. 365--376, 2020.

\bibitem{wardle2017information}
C.~Wardle and H.~Derakhshan, \emph{Information disorder: Toward an
  interdisciplinary framework for research and policymaking}.\hskip 1em plus
  0.5em minus 0.4em\relax Council of Europe Strasbourg, 2017, vol.~27.

\bibitem{nekmat2020nudge}
E.~Nekmat, ``Nudge effect of fact-check alerts: Source influence and media
  skepticism on sharing of news misinformation in social media,'' \emph{Social
  Media+ Society}, vol.~6, no.~1, p. 2056305119897322, 2020.

\bibitem{gb}
Accessed on 2022, \url{https://github.com/compound-finance/}.

\bibitem{op}
Accessed on 2022,
  \url{https://blog.openzeppelin.com/compound-governor-bravo-audit/}.

\end{thebibliography}

 \appendices
\section{Summary of Metrics \& Methods Used}
\label{metric-method}

In this section, we provide a summary along with the current limitations of the metrics and methods used to explore \textbf{RQ2} and \textbf{RQ3}.

\subsection{Metrics to Assess Decentralization}

Decentralization level of DAOs are assessed and compared using four metrics: (1) Interaction Among DAO Holders, (2) DAO Holders' Proposal Pattern, (3) DAO Holders' Voting Power, and (4) Gas Cost to Participate in Voting. For Interaction Among DAO Holders, our study employs a network analysis to identify different clusters of voters and holders based on voting behavior for different proposals. For DAO Holders' Proposal Pattern, logistic regression analysis is conducted to explore the dependency between proposal authors token balances and proposal outcome. In DAO Holders' Voting Power, we explore decentralization by calculating well-known decentralization and inequality indexes (e.g., Gini co-efficient) on voting power distribution and measuring the participation rate between those with different magnitude of voting power. Lastly, in Gas Cost to Participate in Voting was compared across DAOs to understand what aspect of the governance process different DAO smart contracts prioritize and to determine if monetary cost impede user participation.

Due to the transparent nature of blockchain, onchain data, such as proposal metadata and vote metadata, are easily accessible. However, it is also important to note that offchain contextual information such as user identity is challenging to obtain. . Moreover, some analyses in this section rely on statistical tests, which would be more meaningful with larger number of data points. This makes analyses done on smaller DAO less likely to yield statistical significant insights.

\subsection{Metrics to Assess Autonomy}

We aim to examine autonomy within DAOs by analyzing four distinct metrics: (1) Capability of Arbitrary Transaction Execution, (2) Third-Party Dependent Proposals After Execution, (3) Canceled Proposals After Voting Ends, and (4) Execution Time Delay. First, Capability of Arbitrary Transaction Execution section examines the involvement of human in existing strategies used by DAOs to relay proposal consensus to onchain transactions. 
Next, Third-Party Dependent Proposals After Execution metric quantify the number of proposals relying on third party to be successfully executed and aggregate methods DAOs employ to minimize the risk when interacting with third parties. 
The Canceled Proposals After Voting Ends --investigates how well consensus is respected and the role of human intervention with the DAO action through proposals canceled after consensus has been reached. Lastly, Execution Time Delay section measures the time difference between proposal intended execution time to actual execution time.

While most of the proposed metrics can be generalized and applied to majority of the DAOs, many of them require manual calculations. This is especially a problem in DAOs that employ offchain voting as a mapping between offchain consensus and onchain transaction is needed. This highlights the need of transparency tools, tracking proposal timeline, in this field. 
Additionally, quantifying third-party dependent proposals objectively can be challenging for external researchers, and the proposed criteria may not be applicable to all proposals in the space.

\section{More Details about The 10 DAOs Studied}
\label{10_dao}
DAOs often adopt existing governance protocols from predecessors. For example, BitDAO's governance token BIT uses the governance design of Compound's governance token COMP, which is built on the governance bravo contract~\cite{gb}. This is done for flexibility in governance design, as it allows for delegated voting (i.e., allowing another person to vote on one's behalf) and off-chain vote aggregation with the potential for future on-chain governance. Some DAOs use factory contracts, such as dxDAO's DAOStack factory contract
while others use platforms like Juicebox for treasury management and Kleros for bridging off-chain decisions to on-chain action. However, the widespread reuse of popular governance protocols can limit the diversity of governance system.
Furthermore, when DAOs reuse existing sets of smart contracts, they are relying on the security features of that codebase. This approach can have both advantages and limitations. Many DAO governance contracts developed from bigger protocols have already been audited for security issues, such as the Compound Governor Bravo Audit~\cite{op}, which can be beneficial for smaller DAOs that may not have the capital or human resources to develop their own smart contracts and have them audited. On the other hand, if a vulnerability is found in a DAO using similar contracts, an attacker could easily target other DAOs that share the same codebase. Additionally, it can be more challenging for DAOs that use others' code to implement unique features that are specific to their organization. In general, while reusing existing smart contract code can be a practical solution for smaller DAOs, it is important to be aware of the potential risks and limitations involved. It is also essential to consider the unique needs and requirements of a particular DAO when deciding whether to reuse existing code or to develop custom smart contracts.

Each DAO has a specific voting period, ranging from a minimum of two days to seven days. After the voting period ends, DAOs have a timelock period, which can vary. Out of the 10 target DAOs, three have a timelock period, for example, CompoundDAO has a 48-hour timelock period while MolochDAO has a 7-day timelock period. Some DAOs, such as Meta Gamma Delta and MolochDAO, also have strategies to add a grace period  
of 12 days into the timelock. On the other hand, BitDAO, AssangeDAO, Proof of humanity, Bankless DAO, and KrauseHouse do not have any timelock period 
While the timelock period could be implemented in addition to a multisig, the DAOs we studied that uses a multisig dont have a timelock.
In particular, BitDAO, AssangeDAO, Bankless and KrauseHouse use the multisig feature where a safe Multisig Admin manages the treasury and ensures executions. This allows organizations to fully customize how they manage their assets and protocols, with the option of requiring a predefined number of signatures to confirm transactions. Smart contracts can execute actions as long as a predefined number of trusted members agree upon them. Finally, none of the DAOs support votes from governance tokens on another chain, except for BanklessDAO.

\section{Demographics Information of DAO Interview Participants}
In this section, we provide a summary of background of the DAO participants that we interviewed.

\subsection{Participants.} 
\label{demo-formative}
Among the eight participants who took part in the one-to-one interviews, seven were male and one was female. All participants had either been involved or had worked in one or more DAOs, such as MakerDAO, Aave, Meta Gamma Delta, Pleasr DAO, Nouns DAO, PolywrapDAO, LlamaDAO, Aragon, LexDAO, Hydra, SporusDAO, and CaliDAO. Some of them were also engaged in DAO analytics and research, security auditing tools, involving DAOhq, DAO research collective, DAOstar, ConsenSys, Onchain LLC, and Polychain. Moreover, the participants had diverse backgrounds and areas of expertise, ranging from protocol development, software engineering, business operations, project management, finance, to legal expertise (i.e., lawyer). The group session comprised 10 experts from various DAOs, Aragon, BlockScience, Uniswap Foundation, former MakerDAO, Gitcoin, TalentDAO, RnDAO, Metagov, gnosisdao, among others. Of the ten participants, eight were male and two were female. Their areas of expertise included developers, project management, governance and policy, co-founder, researcher, among others. Lastly, the breakout session involved more than 10 participants with expertise in DAO governance from both academia and industries.

\section{Additional Figures \& Tables}

\begin{figure*}[ht]
\begin{subfigure}{0.20\textwidth}
  \includegraphics[width=\linewidth]{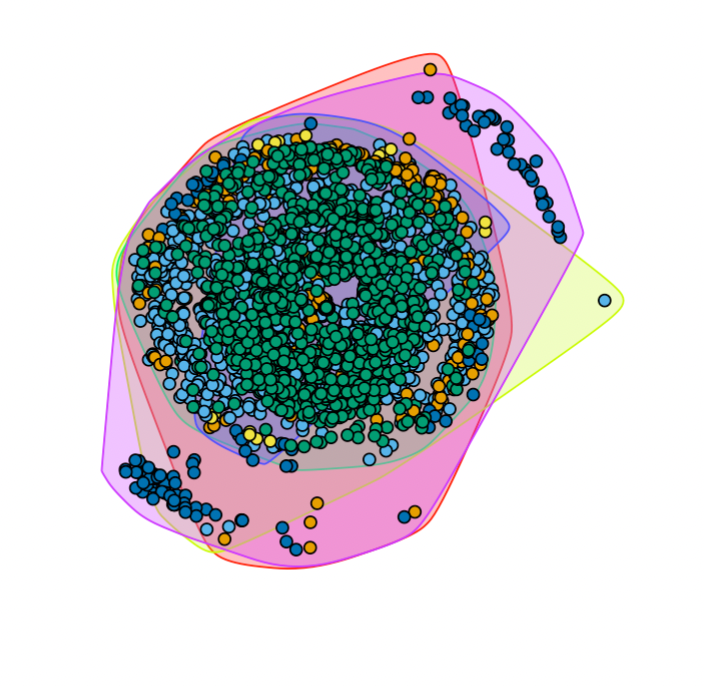}
  \caption{Proof of Humanity DAO}\label{fig:poh_image1}
\end{subfigure}\hfill
\begin{subfigure}{0.20\textwidth}
  \includegraphics[width=\linewidth]{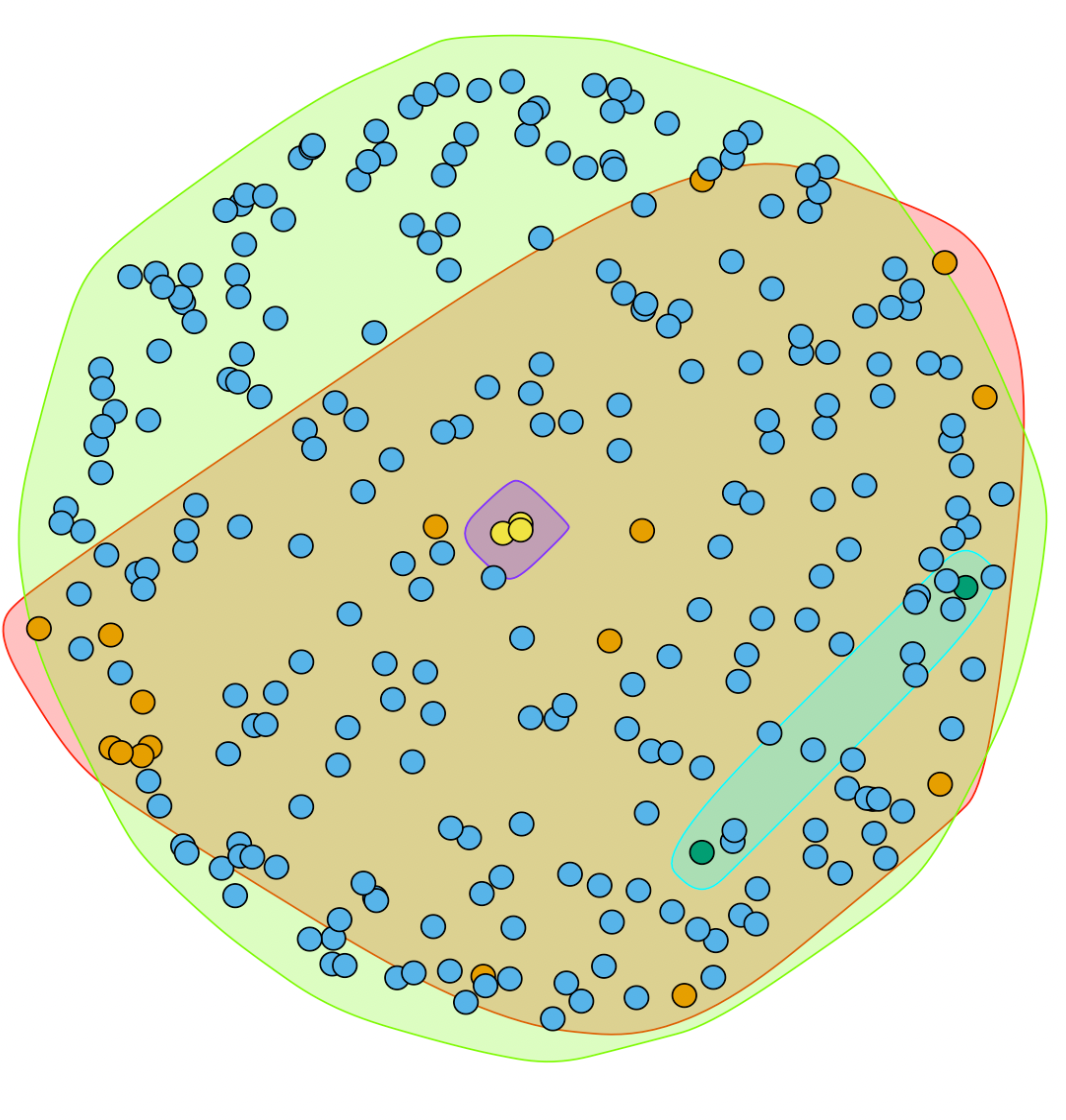}
  \caption{dxDAO}\label{fig:dxdao_image2}
\end{subfigure}\hfill
\begin{subfigure}{0.20\textwidth}%
  \includegraphics[width=\linewidth]{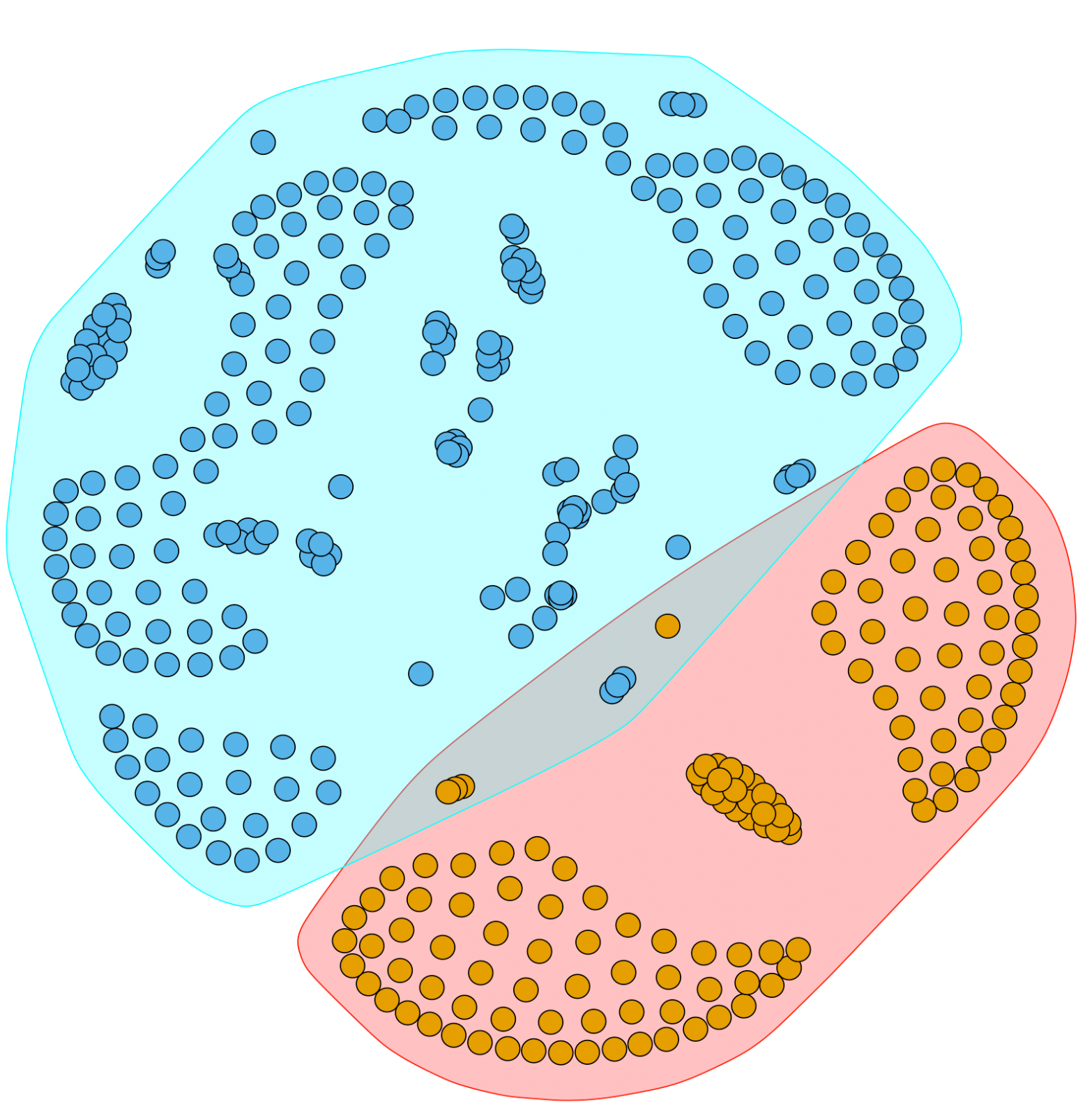}
  \caption{LivePeer DAO}\label{fig:live_image3}
  \end{subfigure}\hfill
\begin{subfigure}{0.20\textwidth}%
  \includegraphics[width=\linewidth]{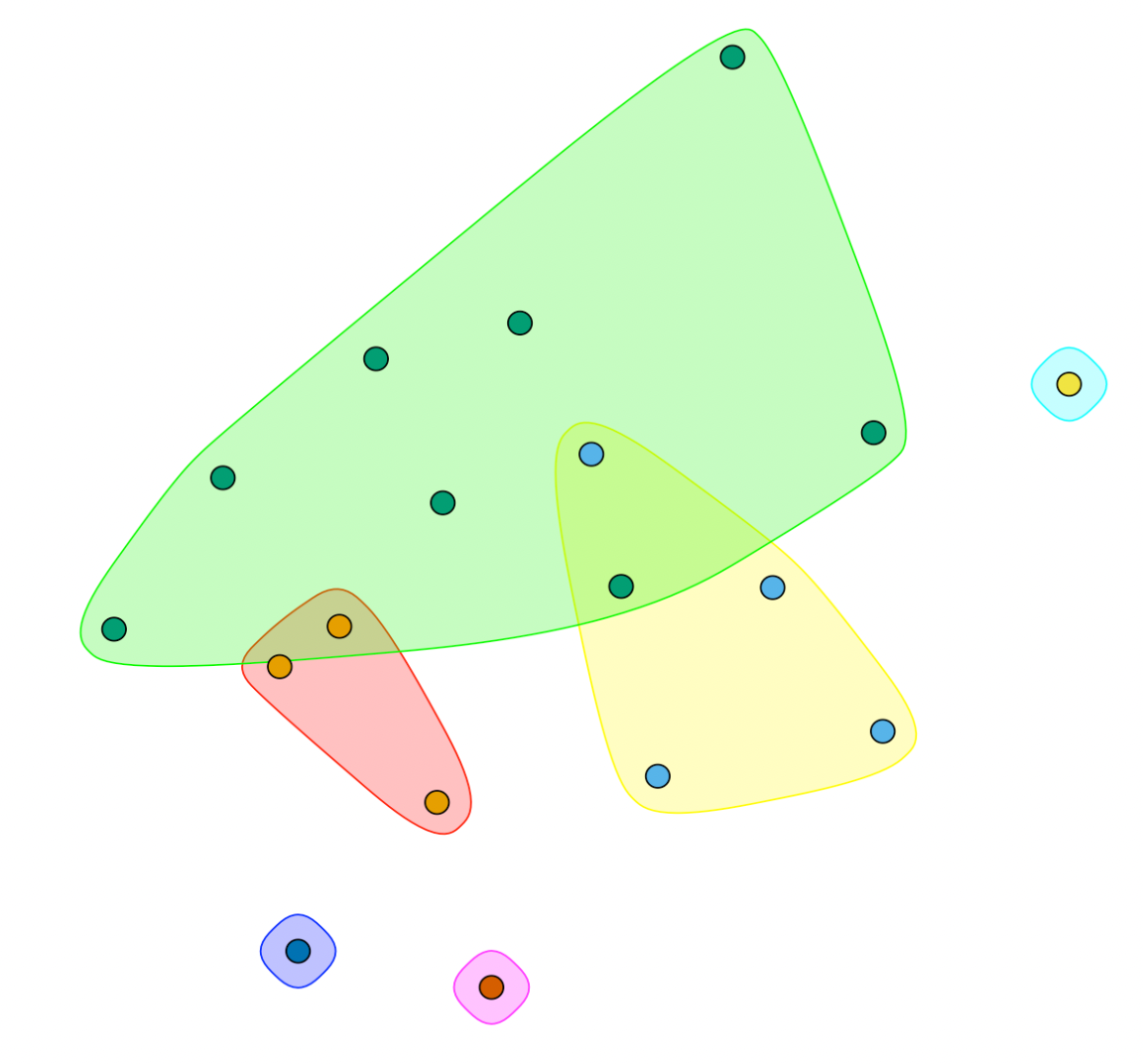}
  \caption{Meta Gamma Delta DAO}\label{fig:mgd_image3}
  \end{subfigure}\hfill
\begin{subfigure}{0.20\textwidth}%
  \includegraphics[width=\linewidth]{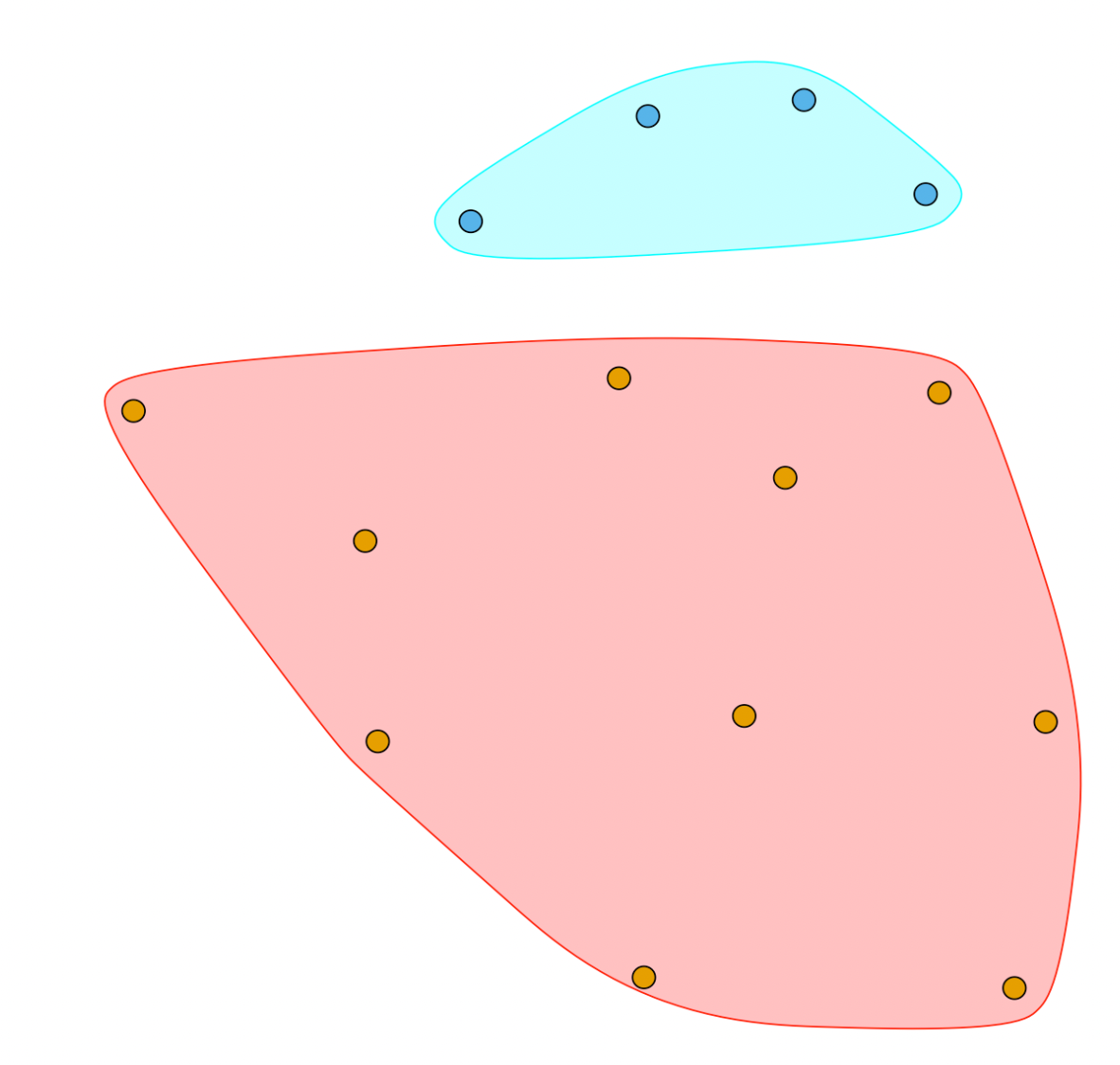}
  \caption{Moloch DAO}\label{fig:moloch_image3}
\end{subfigure}
\caption{DAO holders' voting behavior for different proposals are displayed as community graphs. A cluster is defined based on their voting behavior where 1: voted for certain proposal, -1: voted against certain proposal. The colored polygons in subgraphs represent the nodes with similar voting behavior. There are some overlapped colored regions which indicate that nodes in those clusters have some voting similarity with nodes in other cluster.}
\label{cluster-new}
\end{figure*}

\begin{table}[]
\scriptsize

\begin{tabular}{lllllllllll}
\hline
DAOs        & 
\begin{tabular}[c]{@{}l@{}}Number of\\Whale\end{tabular} &
\begin{tabular}[c]{@{}l@{}}Whale Token \\Range\end{tabular} &
\begin{tabular}[c]{@{}l@{}}Freq of \\ community\end{tabular} &

\\
\hline
Bit& 4.5\% (13/283) & (100m - 1B BIT)&2\\
Assange &3.5\%(38/1066)&(50m-200m Justice)& 2                                \\
PoH&1\%(35/3085)&(2-35 UBI&                 2                            \\  
Bankless&0.3\%(11/3234)&(1m-10m Bank)&       4                               \\
KrauseHouse & 2.3\%(14/592)& (25k-250k Krause)&2\\
Compound    & 1.28\% (32/2482)&(100k-10m COMP)& 2 \\
LivePeer    & 4.3\%(16/368)& (250k-1m)& 2                                \\
MGD         &    NA (same weight) &NA&   NA                                   \\
Moloch      & 46\%(7/15) & (6-100)&   6                                     \\
dxDAO       & 11.8\%(33/278)& NA&2\\                                                     
\hline
\end{tabular}
\caption{Whale addresses number and token holding and number of cluster based on voting similarities}
\label{tab:graph-whale}
\end{table}

\begin{table}[]
\centering
\scriptsize
\begin{tabular}{c|cc}
 \textbf{Name} & \textbf{Entropy}& \textbf{Gini} \\
\hline
CompoundDAO  & -0.0050 (0.001) & 0.0008 (0.015) \\
BitDAO  & -0.0275 (0.490) & 0.0341 (0.000)\\
AssangeDAO & -0.3010 (0.038) & 0.0005 (0.934)\\
 Proof of Humanity & 0.0226 (0.000) & 0.0016 (0.020) \\
BanklessDAO & -0.0568 (0.000) & 0.0067 (0.000) \\
KrauseHouse & -0.0048 (0.014) & 0.0002 (0.507) \\
LivePeer  & -0.0065 (0.903) & 0.0023 (0.571) \\
MGD& 0.0194 (0.000) & 0 \\
MolochDAO  & 0.0125 (0.345) & 0.0063 (0.358)\\
dxDAO & 0.0050 (0.128) & 0.0002 (0.702) 
\end{tabular}
\caption{Decentralization Level Over time. Each value indicates a slope of a linear regression analysis, and the values in parentheses indicate the p-value.}
\label{tab:decen-over}
\end{table}

\begin{table}
    \centering

\small
  \begin{tabular}{l l l l l}
 \hline
DAO & Arbitrary Txn Execution\\ 
 \hline
CompoundDAO &On-chain\\
LivePeer & 2 of 3 Multisig\\
MetaGammaDelta &On-chain\\
dxDAO&On-chain\\
MolochDAO&On-chain\\
AssangeDAO&3 of 7 Multisig\\
BitDAO&3 of 6 Multisig\\
Bankless&4 of 7 Multisig\\
ProofOfHumanity&Kleros Governor\\
KrauseHouse&4 of 7 Multisig\\
 \hline
  \end{tabular}
  \caption{Summary of the arbitrary transaction execution of each DAO. Rows are DAOs. Columns are Arbitrary Transaction Execution Capability and Context note of why that DAO do/doesn’t need it.}

  \label{Tab:Tab-auto-met-1}
\end{table}

\begin{table}[]
\scriptsize
\begin{tabular}{llrrrr}
\hline
DAOs            & Binding                                                        & \multicolumn{1}{l}{\begin{tabular}[c]{@{}l@{}}Total \\ Prop\end{tabular}} & \multicolumn{1}{l}{\begin{tabular}[c]{@{}l@{}}Total \\ Prop \\ Passed\end{tabular}} & \multicolumn{1}{l}{\begin{tabular}[c]{@{}l@{}}Proposal \\ Passed but \\ Did Not Get \\ Executed\end{tabular}} & \multicolumn{1}{l}{\begin{tabular}[c]{@{}l@{}}Total \\ Unknown \\ Status\end{tabular}} \\
\hline
CompoundDAO     & \begin{tabular}[c]{@{}l@{}}Proposer \\ can cancel\end{tabular} & 137                                                                            & 110                                                                                     & 6                                                                                                             & 0                                                                                      \\
dxDAO           & Binding                                                        & 789                                                                            & 678                                                                                     & 0                                                                                                             & 0                                                                                      \\
LivePeer        & Not binding                                                    & 7                                                                              & 7                                                                                       & 0                                                                                                             & 0                                                                                      \\
MolochDAO       & Binding                                                        & 25                                                                             & 25                                                                                      & 0                                                                                                             & 0                                                                                      \\
MetaGammaDelta  & Binding                                                        & 75                                                                             & 69                                                                                      & 0                                                                                                             & 0                                                                                      \\
AssangeDAO      & Not binding                                                    & 11                                                                             & 10                                                                                      & 0                                                                                                             & 0                                                                                      \\
BitDAO          & Not binding                                                    & 18                                                                             & 16                                                                                      & 0                                                                                                             & 2                                                                                      \\
BanklessDAO     & Not binding                                                    & 51                                                                             & 48                                                                                      & 0                                                                                                             & 2                                                                                      \\
ProofOfHumanity & Not binding                                                    & 104                                                                            & 87                                                                                      & 9                                                                                                             & 5                                                                                      \\
KrauseHouse     & Not binding                                                    & 131                                                                            & 113                                                                                     & 14                                                                                                            & 44      \\    
\hline
\end{tabular}
\caption{Proposal decisions in DAOs where are whether the consensus is binding through: smart contract timelock that cant be canceled, smart contract timelock that can be canceled by proposer, multisig, or single wallet.  
}
\label{tab:canceled}
\end{table}

\begin{figure}[!b]
\begin{lstlisting}
 function execTransaction(
        address to,  uint256 value, bytes calldata data, Enum.Operation operation, uint256 safeTxGas, uint256 baseGas, uint256 gasPrice, address gasToken, address payable refundReceiver,  bytes memory signatures
    ) public payable virtual returns (bool success) {
        bytes32 txHash;{
            bytes memory txHashData =
                encodeTransactionData(
                    to, value, data, operation, safeTxGas,
                    baseGas, gasPrice, gasToken,refundReceiver,
                    nonce );
            nonce++;
            txHash = keccak256(txHashData);
            checkSignatures(txHash, txHashData, signatures)  }
\end{lstlisting}
\caption{Multi-sig: Arbitrary txn execution achieves in GnosisSafe Multisig Wallet} 
\label{fig:code-snippet-1}
\end{figure}

\begin{figure}[!b]
\tiny
\begin{lstlisting}
  function execute(bytes32 _proposalId) external {
        MultiCallProposal storage proposal = proposals[_proposalId];
        require(proposal.exist, "must be a live proposal");
        require(proposal.passed, "proposal must passed by voting machine");
        if (schemeConstraints != SchemeConstraints(0)) {
            require(
            schemeConstraints.isAllowedToCall(
            proposal.contractsToCall,
            proposal.callsData,
            proposal.values,
            avatar),
            "call is not allowed");
        }
\end{lstlisting}
\caption{On-chain DAOs: dxDAO Arbitrary txn execution achieved by the MultiCallScheme}
\label{fig:code-snippet-2}
\end{figure}

\end{document}